\documentclass[aps,reprint,amsmath,amssymb,groupaddress,superscriptaddress,pra]{revtex4-2}
\usepackage{graphicx}
\usepackage[utf8]{inputenc}
\usepackage{amsmath}
\usepackage{subfigure}
\usepackage{amsfonts}
\usepackage{amssymb}
\usepackage{bm}
\usepackage{dcolumn}
\usepackage{array}  
\usepackage{diagbox}
\usepackage{color}
\usepackage[utf8]{inputenc}
\usepackage[T1]{fontenc}
\usepackage{booktabs, array, mathptmx, float, tabularx, booktabs, lipsum, amsmath, multirow,mathtools}
\usepackage{soul}
\usepackage{xcolor}
\usepackage{mathtools}
\usepackage{lipsum}
\usepackage{comment}

\usepackage{xcolor}
\usepackage[colorlinks,citecolor=blue,linkcolor=blue,anchorcolor=blue,urlcolor=blue]{hyperref}
\begin{document}
\title{Unified approach to power-efficiency trade-off relations of generic thermal machines}

\author{Yu-Han Ma}
\email{yhma@bnu.edu.cn}
\affiliation{School of Physics and Astronomy, Beijing Normal University, Beijing, 100875, China}
\affiliation{Key Laboratory of Multiscale Spin Physics (Ministry of Education), Beijing Normal University, Beijing 100875, China}
\author{Cong Fu}
\email{fucong@stu.xmu.edu.cn}
\affiliation{School of Physics and Astronomy, Beijing Normal University, Beijing, 100875, China}
\affiliation{Department of Physics, Xiamen University, Xiamen 361005, Fujian, China}

\date{\today}
\begin{abstract}
We present a general framework for determining the power-efficiency trade-off relations across arbitrary thermal machines, addressing the lack of unified optimization results stemming from their diverse functionalities (e.g., heat engines, refrigerators, and heat pumps). For time-dependent cycle irreversibility $A(\tau)$ following a $\tau^{-\alpha}$ power law, where $\alpha$ is an interaction-dependent parameter, we show that engineering the interactions between thermal machines and reservoirs enables control over the trade-off relations, with the efficiency at maximum power approaching Carnot efficiency as $\alpha$ increases. Setting $\alpha=1$ naturally recovers typical low-dissipation regime results. Additionally, we derive the first power-efficiency trade-off for finite-time quantum adiabatic Otto machines with $\tau^{-2}$-scaling. This work establishes a unified constraint for thermodynamic cycles across non-equilibrium regimes, facilitating consistent optimization of diverse thermal devices in practice.

\end{abstract}
\maketitle

\section{Introduction}
\label{I}
Finite-time thermodynamics bridges the gap between idealized quasi-static cycles in equilibrium thermodynamics and real-world thermal machines operating within finite cycle times, standing as a widely employed non-equilibrium thermodynamic theoretical approach of substantial practical importance~\cite{andresenThermodynamicsFiniteTime1977a,salamonThermodynamicsFiniteTime1977,ondrechenThermodynamicsFiniteTime1980}. A central challenge in this field is characterizing the performance of thermal machines quantitatively through the power-efficiency trade-off relation (hereinafter referred to as the trade-off relation for brevity)~\cite{qiu2025roadmap,chenEffectHeatTransfer1989, holubecEfficiencyMaximumPower2015,shiraishi2016universal, holubecMaximumEfficiencyLowdissipation2016, maUniversalConstraintEfficiency2018, pietzonkaUniversalTradeOffPower2018, yuanOptimizingThermodynamicCycles2022,zhou2024finite,zhao2025revisiting}, which arises from energy dissipation due to irreversibilities~\cite{shiraishi2016universal, brandnerThermodynamicGeometryMicroscopic2020,maExperimentalTestTau2020,zhaiExperimentalTestPowerefficiency2023}. Previous studies have derived trade-off relations for specific models, such as the endo-reversible Carnot heat engine~\cite{chenEffectHeatTransfer1989,zhao2025revisiting}, linear irreversible heat engines~\cite{holubecEfficiencyMaximumPower2015,yuanOptimizingThermodynamicCycles2022}, and low-dissipation heat engines~\cite{holubecMaximumEfficiencyLowdissipation2016,maUniversalConstraintEfficiency2018}. Beyond these conventional trade-off relations, the thermodynamic trade-offs in diverse practical scenarios, including information-energy conversion~\cite{maMinimalEnergyCost2022,zhou2024finite}, particle separation~\cite{chen2023geodesic,zhaoEngineeringRatchetbasedParticle2024}, battery charging~\cite{lei2025universal}, and biochemical motors~\cite{zhai2025power}, have recently garnered growing research interest. Besides, one key parameter of finite-time thermal machines, efficiency at maximum power (EMP)~\cite{qiu2025roadmap,curzonEfficiencyCarnotEngine1975,tuEfficiencyMaximumPower2008,vandenbroeckThermodynamicEfficiencyMaximum2005,espositoEfficiencyMaximumPower2010,blanchardCoefficientPerformanceFinite1980}, stands out as a typical point on the trade-off relation~\cite{maUniversalConstraintEfficiency2018,Chen1996TheIO,wangCoefficientPerformanceMaximum2012,izumidaCoefficientPerformanceOptimized2013}, delineating the energy conversion efficiency when the power is maximized. 

However, research on the trade-off relation in thermal machines still suffers from two critical limitations. The first lies in the lack of universality across thermal machine types. Existing trade-off relations are predominately derived for heat engines, while investigations into refrigerators and heat pumps remain notably insufficient~\cite{holubecMaximumEfficiencyLowdissipation2020,yeMaximumEfficiencyLowdissipation2022}. More critically, no universal trade-off relation has been established to unify different thermal machine categories. The core obstacle stems from the absence of a cohesive framework capable of describing these diverse machines—rooted in the inconsistency of their performance metrics~\cite{chenStudyOptimalPerformance1995,de2013low,johalPerformanceOptimizationLowdissipation2019,gonzalez-ayalaEntropyGenerationUnified2018}. For instance, energy conversion efficiency, a key parameter for characterizing trade-offs, is defined distinctly across devices: heat engines rely on "efficiency", while refrigerators and heat pumps use the "coefficient of performance (COP)"~\cite{leffEERCOPSecond1978,gordonOptimizingChillerOperation1997}. This divergence in metrics makes a one-size-fits-all trade-off relation for general thermodynamic cycles seemingly unachievable, forcing optimization to be conducted on a case-by-case basis for different thermal machines. The second limitation concerns the restricted applicable regime of the trade-off relations. Due to the equivalence among three core thermal machine models—endoreversible, linear irreversible, and low-dissipation~\cite{wang2012efficiency,johal2017heat}, nearly all existing trade-off relations~\cite{chenEffectHeatTransfer1989,holubecEfficiencyMaximumPower2015,holubecMaximumEfficiencyLowdissipation2016,maUniversalConstraintEfficiency2018,yuanOptimizingThermodynamicCycles2022,zhaiExperimentalTestPowerefficiency2023,zhao2025revisiting} are tied to the typical $1/\tau$-irreversibility ($\tau$ denotes cycle duration)~\cite{espositoUniversalityEfficiencyMaximum2009,maUniversalConstraintEfficiency2018,maExperimentalTestTau2020}. Nevertheless, recent studies have revealed that the intrinsic specificity of working substances~\cite{chen2019achieve,fei2022efficiency,liang2023} or the diversity of system-reservoir interactions~\cite{pancotti2020speed} can modify the low-dissipation characteristics of heat engines, thereby significantly affecting their finite-time performance. Consequently, developing a universal trade-off relation that transcends the $1/\tau$-irreversibility regime is urgently needed to optimize practical thermal devices in broader, real-world operating regimes.

In this work, we tackle the aforementioned challenges by deriving a unified trade-off relation: universal for all thermal machine (any function) and beyond the low-dissipation regime. By connecting the $\tau$-dependent irreversibility $A(\tau)\propto\tau^{-\alpha}$~\cite{yang2013bounds,pancotti2020speed} to the performance parameters of thermal machines, we succinctly derive the trade-off relation for a generic thermal machine. 
Our general result naturally recovers the previous methods when the dimensionless parameter $\alpha$, which depends on the system–reservoir interaction~\cite{pancotti2020speed}, is specified, as will be shown later. The concrete contents of this paper are organized as follows: In Sec.~\ref{II}, we develop a mathematical framework for the general thermal machine, where the sign functions are introduced and the generalized efficiency and dissipation are defined. In Sec.~\ref{IIIA}, we derive the trade-off relation between power and efficiency from the unified description of thermal machines, together with the lower bound of dissipation. In Sec.~\ref{IIIB}, we further investigate the EMP of the thermal machines. Two typical non-equilibrium operating regime of finite-time thermal machines are discussed in detail in Sec.~\ref{IV} to demonstrate the validity of our general results. Finally, we draw our conclusions in Sec.~\ref{V}.

\section{Unified description of thermal machines}
\label{II}
In this section we employ the sign function as a powerful tool to characterize the efficiency of the thermal machines. By decomposing the system energy into reversible and irreversible contributions, we derive a unified relation between efficiency and dissipation.

\begin{table}[h]
\caption{\label{tab:table1}%
Cycle quantities of thermal machines
}
\begin{ruledtabular}
\begin{tabular}{ccccccc}
&$\theta$&
$\gamma$&
$\rm{O}$&
$\rm{I}$&
$\eta$&
$P$\\
\colrule
Heat Engine & $+1$& $+1$ & $W$ & $Q^{\rm{h}}$& $W/Q^{\rm{h}}$& $W/\tau$\\
Heat Exchanger & $+1$ & $-1$& $Q^{\rm{c}}$ & $Q^{\rm{h}}$& $Q^{\rm{c}}/Q^{\rm{h}}$& $Q^{\rm{c}}/\tau$\\
Refrigerator & $-1$& $-1$ & $Q^{\rm{c}}$ & $W$&$Q^{\rm{c}}/W$& $Q^{\rm{c}}/\tau$\\
Heat Pump & $-1$ & $+1$& $Q^{\rm{h}}$ & $W$& $Q^{\rm{h}}/W$&$Q^{\rm{h}}/\tau$\\
\end{tabular}
\end{ruledtabular}
\end{table}

\begin{figure}[h]
        \centering		\includegraphics[width=0.48\textwidth]{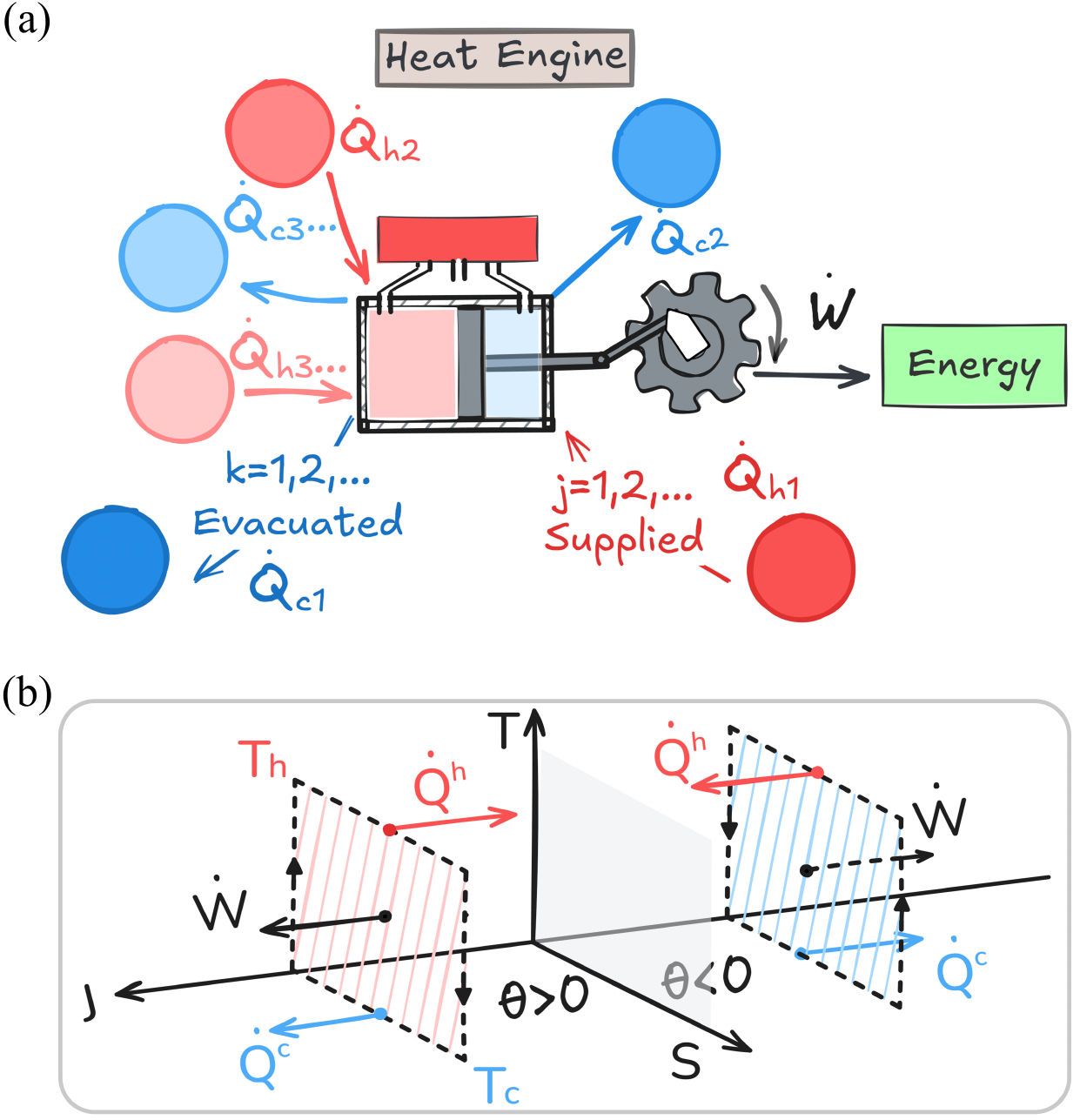}
	\caption{(a) schematic of a heat engine operates among multiple heat reservoirs which is represented by red and blue circles. The engine extracts from supplied reservoirs and releases to others. (b) Temperature-entropy ($T-S$) diagram of clockwise and counterclockwise finite-time Carnot cycles. The orientation of the $J$-axis represents the direction of energy flow from the cycles to the heat reservoirs.}
\label{fig:multiple_bath}
\end{figure}

Thermal machines are designed to convert energy in various forms according to actual demands, including heat engines, heat pumps, refrigerators, and heat exchangers. Although heat exchangers are typically applied to individual heat transfer processes rather than full thermodynamic cycles in engineering practice, we include them in Table~\ref{tab:table1} for completeness. Among them, heat engines harness the temperature difference between hot and cold heat reservoirs to perform work. Both refrigerators and heat pumps create heat flow from cold reservoirs to hot reservoirs through work input, one for cooling and the other for heating. In general sense, the energy conversion efficiency of any thermal machine can be broadly cognitively defined as $\eta \equiv \rm{O}/\rm{I}$, where $\rm{O}$ and $\rm{I}$ are respectively the energy exchange at the input side and output side of the machine, as illustrated in Tab.~\ref{tab:table1}. It should be mentioned here that "output" is categorized from the perspective of the machine's function, not necessarily the direction of energy flow. For instance, for a general heat engine cycle illustrated in Fig.~\ref{fig:multiple_bath}(a), its output function is performing work externally, $\rm{O}$ represents work output $W$ and $\rm{I}$ represents total heat absorption $Q^{\rm{h}}=\sum_{j}Q^{{\rm{h}},j}$ with $Q^{{\rm{h}},j}$ the heat absorption from the $j$-th hot reservoir (We do not restrict the number $N$ of reservoirs, and the following discussion is applicable to any situation with $N\geq2$)~\cite{amelkinMaximumPowerProcesses2004}. In this case, the direction of energy flow at the output side of the machine is also from the machine to the external environment. However, for a refrigerator cycle, where the output function is absorbing heat from the low-temperature side, namely the direction of energy flow is from the surroundings to the refrigerator. 

For thermal machines operating within finite cycle time, by dividing $\rm{O}$($\rm{I}$) into its scalar-valued reversible part $\rm{O}_{\rm{rev}}$($\rm{I}_{\rm{rev}}$) and irreversible part $A_{\rm{o}}$($A_{\rm{i}}$) on the basis of the second law of thermodynamics, we further express the machine efficiency as
\begin{equation}
\eta=\frac{\rm{O}_{\rm{rev}}-\theta\gamma A_{\rm{o}}}{\rm{I}_{\rm{rev}}-\theta A_{\rm{i}}}.
\label{eq:efficiency}
\end{equation}
Here, $\rm{O}_{\rm{rev}}$, $\rm{I}_{\rm{rev}}$, and $A_{\rm{o}/\rm{i}}$ are all positive. The efficiency expression presented above stems from the following physical intuition: $\theta=\pm1$ is a sign function determined by the direction of the cycle, while $\gamma=\pm1$ is another sign function determined by the specific function of the machine output with a given $\theta$, as illustrated in Tab.~\ref{tab:table1}.

Specifically, after determining the cycle direction $\theta$, the direction of heat flow is also determined, either from the high- to the low-temperature heat reservoir or vice versa. For such a three-side machine (comprising two heat sides and one work side), we may define the side where heat flow converges as the net sink. If the net sink aligns with the machine's desired output side, the condition $\theta \gamma < 0$ holds. This also leads to monotonic behavior in the optimization curve of the heat pump. For example, in heat exchanger and heat pump where $\theta \gamma <0$ , the output cold reservoir(or hot reservoir for heat pump) is just defined at the net sink of the energy flow path; in contrast, for heat engine and refrigerator, the output corresponds to the intermediate tap terminal of the energy flow, respectively. Quantitatively, in Fig.~\ref{fig:multiple_bath}(b), a clockwise ($\theta>0$) and a counterclockwise ($\theta<0$) finite-time Carnot cycle between two reservoirs of temperatures $T_{\rm{h}}>T_{\rm{c}}$ is depicted, where $J$-axis indicates the direction of energy flow exiting the cycles. In such a geometric representation, the greater the output power of the cycle, the further away it is from the plane of $J=0$.
\begin{table}[h]
\caption{\label{tab:table2}
Dissipation and its ratio associated with input and output sides of thermal machines
}
\begin{ruledtabular}
\begin{tabular}{cccccc}
&$A_{\rm{o}}$&
$A_{\rm{i}}$&
$A$&
$A_{\rm{o}}/A$&
$A_{\rm{i}}/A$\\
\colrule
Heat Engine & $W_{\rm{ir}}$& $Q_{\rm{ir}}^{\rm{h}}$ &$W_{\rm{ir}}$&$1$ & $\delta$\\
Heat Exchanger & $Q_{\rm{ir}}^{\rm{c}}$&$Q_{\rm{ir}}^{\rm{h}}$ &$W_{\rm{ir}}$&$1-\delta$&$\delta$\\
Refrigerator & $Q_{\rm{ir}}^{\rm{c}}$& $W_{\rm{ir}}$ & $W_{\rm{ir}}$ & $1-\delta$&$1$\\
Heat Pump & $Q_{\rm{ir}}^{\rm{h}}$ & $W_{\rm{ir}}$& $W_{\rm{ir}}$& $\delta$&$1$\\
\end{tabular}
\end{ruledtabular}
\end{table}

The dissipation $W_{\rm{ir}}$ in the work output/input process, denoted as $A$, is contributed by all heat transfer processes together. This additivity of dissipation is jointly induced by the first and second laws of thermodynamics. Taking a heat engine cycle as an example, we have $Q^{\rm{h}}-Q^{\rm{c}}=W$ and $Q^{\rm{h}}_{\rm{rev}}-Q^{\rm{c}}_{\rm{rev}}=W_{\rm{rev}}$. Subtract the above two equations yields $\left(Q^{\rm{h}}_{\rm{rev}}-Q^{\rm{h}} \right)+\left(Q^{\rm{c}}-Q^{\rm{c}}_{\rm{rev}} \right)=W_{\rm{rev}}-W$, which can be further simplified as $Q^{\rm{h}}_{\rm{ir}}+Q^{\rm{c}}_{\rm{ir}}=W_{\rm{ir}}$. For thermal machines, we illustrate the dissipation in different processes in Tab.~\ref{tab:table2}, where the dissipation ratios of different processes are determined with the definition $\delta\equiv Q_{\rm{ir}}^{\rm{h}}/W_{\rm{ir}}\in[0,1]$. In actual implementations, $\delta$ can be tuned via the thermal properties of system-reservoir interaction strength. For an ideal gas, we derive $\delta$ as a function of system parameters in Appendix~\ref{appendixA} via thermodynamic geometry approach. Taking use of the cycle parameters, we can unify output dissipation $A_{\rm{o}}$ and input dissipation $A_{\rm{i}}$ in terms of $A$ as
\begin{equation}
A_{\rm{i}}=\delta^{\frac{1+\theta}{2}}A,~~ A_{\rm{o}}=\left(1-\delta\right)^{\frac{1-\gamma}{2}} \delta^{\frac{(1-\theta)(1+\gamma)}{4}}A.
\label{A}
\end{equation}

\section{Universal thermodynamic bounds}
\label{III}
In this section, within the theoretical framework developed above for thermal machines of different functionalities and under the assumption of a lower bound on dissipation, we derive the trade-off relations and the EMPs. We then analyze the influence of the parameter $\alpha$ on this relation and illustrate the results with examples of a heat engine and a refrigerator. Notably, for sufficiently large $\alpha$ there exists a parameter regime in which operation at maximum power can coincide with efficiency approaching the reversible limit.
\subsection{Trade-off relations between power and efficiency}
\label{IIIA}
Substituting Eq.~\eqref{A} into Eq.~\eqref{eq:efficiency} yields 
\begin{equation}
\eta=\frac{{\rm{O}}_{\rm{rev}}-\theta\gamma\left( 1-\delta\right)^{\frac{1-\gamma}{2}} \delta^{\frac{(1-\theta)(1+\gamma)}{4}}A}{{\rm{I}}_{\rm{rev}}-\theta \delta^{\frac{1+\theta}{2}}A},
\label{eq:efficiency1}
\end{equation}
such that we can rewrite $A$ in term of $\eta$ as
\begin{equation}
A=\frac{\chi\left(\rm{O}_{\rm{rev}}-\rm{I}_{\rm{rev}}\eta\right)}{\theta\left(1-\chi\eta \right)\delta^{\frac{1+\theta}{2}}},
\label{eq:dissipation}
\end{equation}
where
\begin{equation}
\chi\equiv \gamma\left(1-\delta\right)^{\frac{\gamma-1}{2}} \delta^{\left[\theta-\frac{(1-\theta)(\gamma-1)}{4}\right]}
\label{chi}
\end{equation}
Physically, it is reasonable to assume that $A$ has a cycle duration $\tau$-dependent lower bound~\cite{salamonThermodynamicLengthDissipated1983,andresenThermodynamicGeometryMetrics1988,hoffmannMeasuresDissipation1989,crooksMeasuringThermodynamicLength2007a,chenBoostingPerformanceQuantum2019,maExperimentalTestTau2020,liGeodesicPathMinimal2022,frimGeometricBoundEfficiency2022}, which has been explored in theoretical and experimental aspects i.e., $A\geq A_{\rm{min}}(\tau)$. By further noticing that the output power of the machine is $P=\rm{O}/\tau$, Eq.~\eqref{eq:dissipation} yields
\begin{equation}
\frac{\chi\left(\rm{O}_{\rm{rev}}-\rm{I}_{\rm{rev}}\eta\right)}{\theta\left(1-\chi\eta \right)\delta^{\frac{1+\theta}{2}}}\geq A_{\rm{min}}(\tau)=A_{\rm{min}}\left(\frac{\rm{O}}{P}\right).
\label{eq:dissipationA}
\end{equation}
Eliminating ${\rm{O}}=\rm{O}_{\rm{rev}}-\theta\gamma A_{\rm{o}}$ from the right side of the above equation with Eqs.~\eqref{A} and \eqref{eq:dissipation}, a general inequality with respect to $P$ and $\eta$ is obtained.

\begin{figure}[H]	\includegraphics[width=0.48\textwidth]{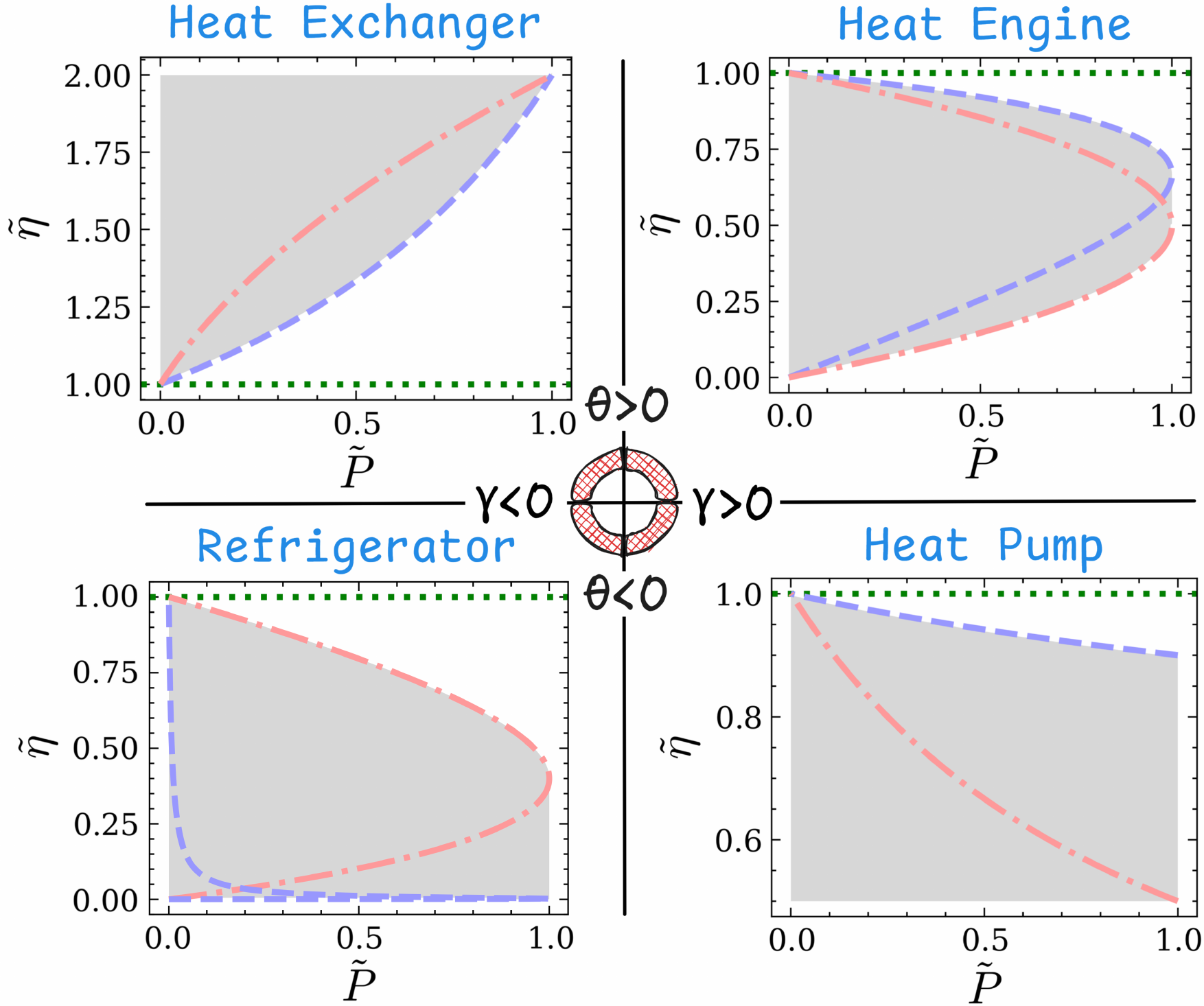}
    \caption{The normalized power-efficiency trade-off relation ($\tilde{\eta}\equiv \eta/\eta_{\mathrm{rev}},~\tilde{P}\equiv P/P_{\mathrm{max}}$) of four thermal machines. The blue dashed (red dash-dotted) curve represents the Eq.~\eqref{eq:tradeoff_general} with $\delta\rightarrow1$($\delta\rightarrow0$), the normalized reversible efficiency $\Tilde{\eta}=1$ is plotted with the green dotted line, and the shadow area marks the available operation region of the machines. In this plot, $\alpha=1$ and $\eta_{\rm{rev}}=0.5$ are used.}
	\label{fig:alpha=1_tradeoff}
    \end{figure}

As a concrete yet general case, we consider $A_{\rm{min}}=\Sigma \tau^{-\alpha}$ with $\Sigma$ denoting a dissipation coefficient~\cite{yang2013bounds,chen2019achieve,pancotti2020speed}. This form captures a wide range of scenarios, including finite-time isothermal processes in the slow-driving regime~\cite{espositoEfficiencyMaximumPower2010,maExperimentalTestTau2020} and finite-time quantum adiabatic processes with energy oscillations neglected~\cite{chen2019achieve,chenBoostingPerformanceQuantum2019}. In this case, Eq.~\eqref{eq:dissipationA} becomes
\begin{equation}
\frac{\chi\left(\rm{O}_{\rm{rev}}-\rm{I}_{\rm{rev}}\eta\right)}{\theta\left(1-\chi\eta \right)\delta^{\frac{1+\theta}{2}}}\geq \Sigma \left( \frac{P}{\rm{O}} \right)^{\alpha}=\Sigma \frac{P^\alpha}{(\rm{O}_{\rm{rev}}-\theta\gamma A_{\rm{o}})^\alpha},
\label{eq:A_L_P_1_gene}
\end{equation}
substituting Eqs.~\eqref{A} and \eqref{eq:dissipation} into which and performing straightforward calculation yields the main result of this work
\begin{equation}
P \le \left( \frac{\mathrm{I}_{\mathrm{rev}}^{\alpha +1} \chi}{\Sigma \theta \delta^{\frac{1+\theta}{2}}} \right)^{\frac{1}{\alpha}} \frac{\eta \, (\eta_{\mathrm{rev}} - \eta)^{\frac{1}{\alpha}} \, (1 - \chi \eta_{\mathrm{rev}})}{(1 - \chi \eta)^{\frac{\alpha +1}{\alpha}}} \equiv P_m(\eta),
\label{eq:tradeoff_general}
\end{equation}
where $\eta_{\rm{rev}}=\rm{O}_{\rm{rev}}/\rm{I}_{\rm{rev}}$ is the reversible efficiency. The equal sign in this unified power-efficiency trade-off relation is saturated when the cycle dissipation is minimized. Figure~\ref{fig:alpha=1_tradeoff} intuitively demonstrate the generality of Eq.~\eqref{eq:tradeoff_general} across different thermal machines. This example ($\alpha=1$) will be discussed in detail in Sec.~\ref{IVA}.

\begin{figure}[h]
        \centering		\includegraphics[width=0.48\textwidth]{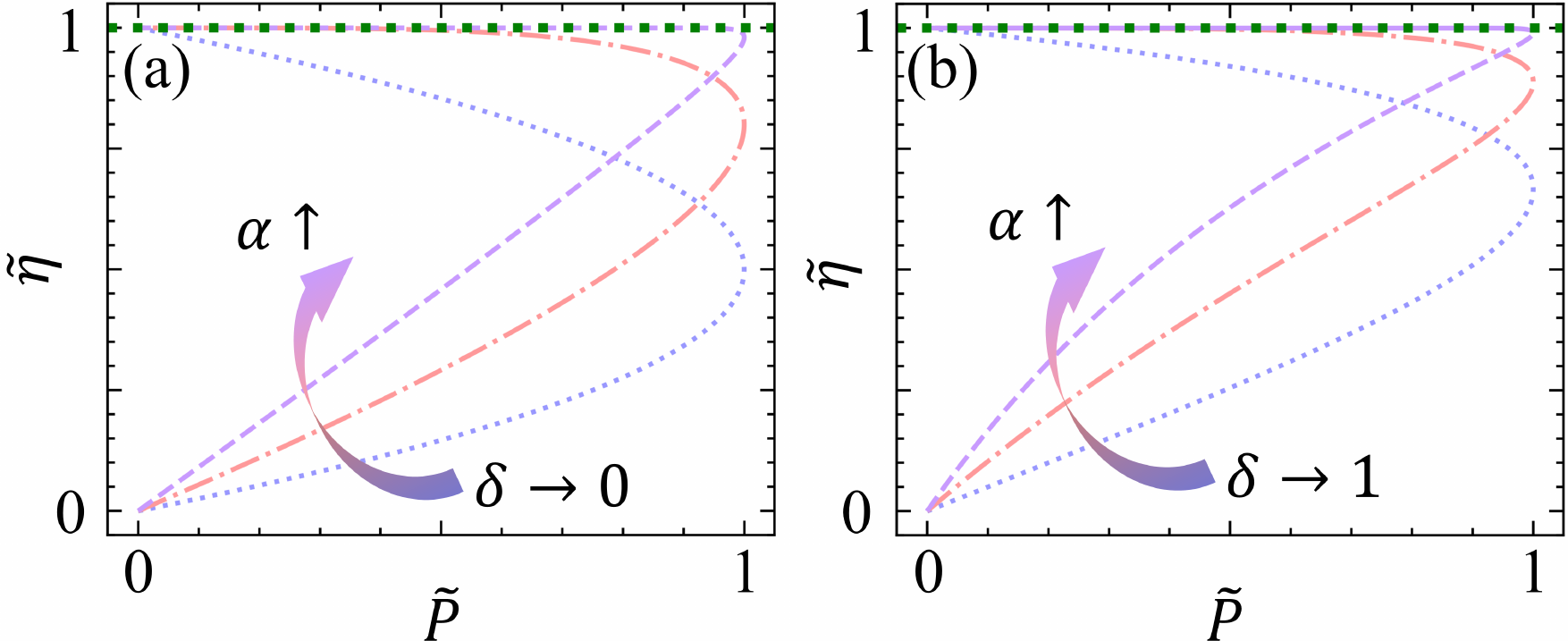}
	\caption{The normalized trade-off relations of heat engines associated with $\delta \to 0$ [(a)] and $\delta \to 1$ [(b)] with different $\alpha$ and $\eta_{\rm{rev}}=0.5$. The blue dotted, red dash-dotted, and purple dashed curve represent $\alpha=1,4,50$, respectively, $\Tilde{\eta}=1$ is plotted with the green dotted line.}
\label{fig:tradeoff_alpha_engine}
\end{figure}

Furthermore, it follows from Eq.~\eqref{eq:tradeoff_general} that the normalized power-efficiency trade-off relations for heat engines ($\theta=\gamma=1$) are illustrated in Fig.~\ref{fig:tradeoff_alpha_engine} for two limiting cases of $\delta=0,1$. As shown in this figure, when $\alpha$ increases, the extreme point on the right-hand side of the trade-off relation gradually approaches $(1,1)$, indicating that the heat engines can approach reversible efficiency at maximum power. This can be regarded as a thermodynamic advantage~\cite{liang2023}. We will further discuss the properties of the EMP in Sec.~\ref{IIIB}. Notably, the dissipation $A\propto\tau^{-\alpha}$ can be achieved by smoothly modulating the system-reservoir interaction, where any $\alpha > 0$ can be attained through appropriate modifications of the interaction~\cite{pancotti2020speed}. Particularly, $\alpha=1$ corresponds to a standard isothermal process with constant system-reservoir interaction~\cite{maUniversalConstraintEfficiency2018,pancotti2020speed,maExperimentalTestTau2020}. The parameter $\alpha$ is introduced not only to generalize the expression of our unified approach, but also to provide an additional degree of freedom for optimizing the performance of thermal machines. This degree of freedom offers a more general perspective for enhancing thermodynamic processes and enables broader control over the operational characteristics of such systems. 

\begin{figure}[h]
        \centering		\includegraphics[width=0.48\textwidth]{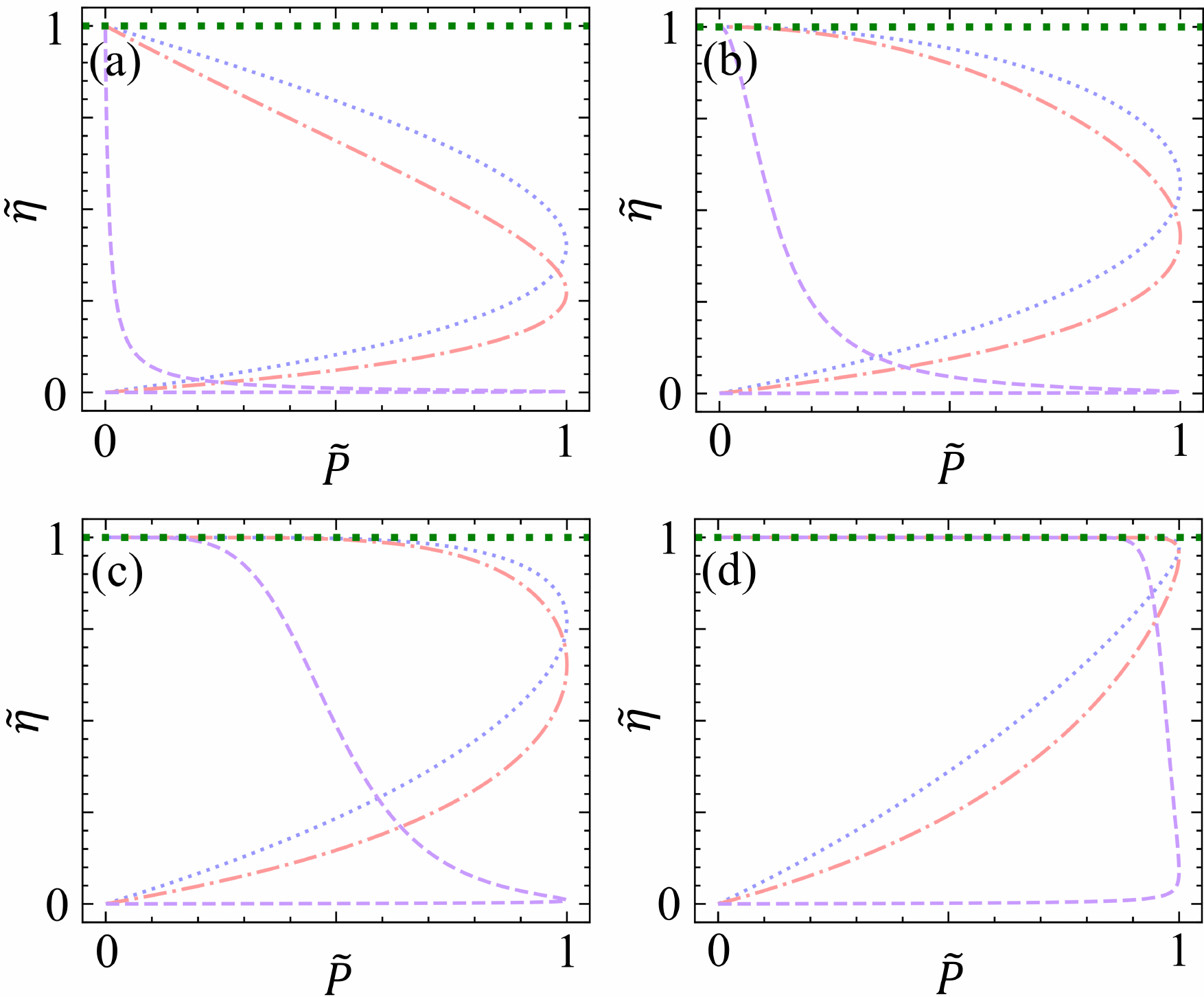}
	\caption{The normalized trade-off relations associated with $\alpha = 1$[(a)], $\alpha = 2$[(b)], $\alpha = 5$[(c)], and $\alpha = 50$[(d)]. The blue dotted, red dash-dotted, and purple dashed curves are plotted with $\delta=0,0.7,1$, respectively. In this plot, $\eta_{\rm{rev}}=0.5$.}
\label{fig:refri_alpha}
\end{figure}

By specifying parameter $\theta=\gamma=-1$, we demonstrate the trade-off relations of refrigerators in Fig.~\ref{fig:refri_alpha} with different $\delta$ and $\alpha$. As $\delta$ increases, the maximum power point on the right-hand side of the trade-off relation moves downward and gradually approaches (1,0). This differs from the $\delta$-monotonicity observed in heat engines, owing to the distinct definition of efficiency. Moreover, when $\delta \nrightarrow 1$, increasing $\alpha$ causes the entire trade-off relation to shift upward, similar to the behavior of heat engines. Specifically, in the Fig.~\ref{fig:refri_alpha}(d), for large $\alpha$ and $\delta=1$, the trade-off forms a trapezoid--a behavior not previously observed. This indicates that the selection of EMP appears to be arbitrary. In the $\eta-P$ diagram, the long-time regime occupies the upper section of the curve, while the short-time regime resides in the lower section. A critical time threshold exists within the short-time regime: below this threshold, power increases sharply, but refrigerating performance remains low and exhibits strong time dependence. Above the threshold, power flattens as irreversibility diminishes significantly with increasing duration, driving substantial efficiency enhancements.
After examining the derivative formula of $P_m(\eta)$ given in Eq.~\eqref{eq:tradeoff_general}, it was observed that when the efficiency approaches zero or reaches the reversible efficiency $\eta_{\mathrm{rev}}$, both the initial value of the derivative and its rate of change become significantly large. This phenomenon causes the trade-off relation curve between $P$ and $\eta$ to rapidly decline to zero near these two extreme points. In contrast, in the middle region of the efficiency regime, the derivative remains relatively small and nearly constant, resulting in minimal variation in $P$ with respect to $\eta$. Consequently, a trapezoid pattern emerges in the graphical representation.

\subsection{Efficiency at maximum power}
\label{IIIB}
\begin{figure}[h]
        \centering		\includegraphics[width=0.48\textwidth]{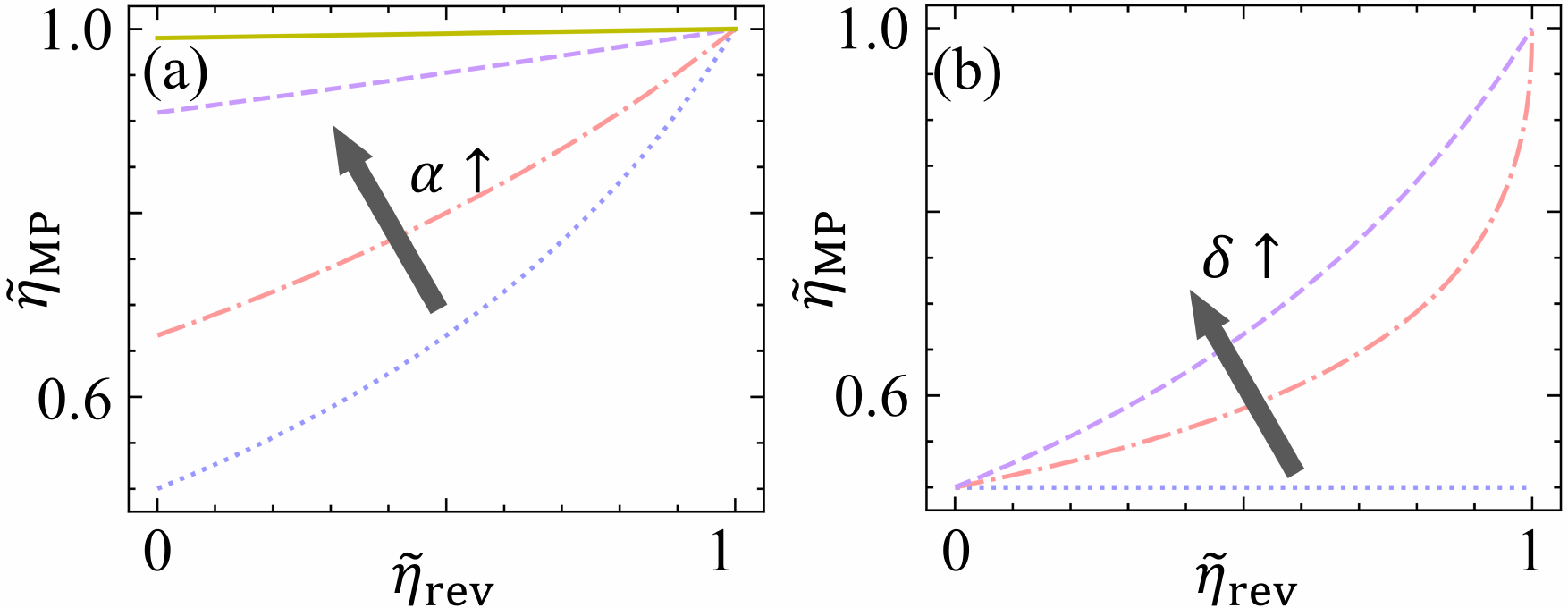}
	\caption{Normalized EMPs of heat engines ($\Tilde{\eta}_{\rm{MP}} \equiv \eta_{\mathrm{MP}}/\eta_{\mathrm{rev}}$) as functions of $\tilde{\eta}_{\rm{rev}}$. (a) The blue dotted, red dash-dotted, yellow solid, and purple dashed curves are plotted with $\alpha=1,2,10,100$, respectively, and $\delta=1$ is fixed. (b) The blue dotted, red dash-dotted, and purple dashed curves are associated with $\delta=0,1/(1+\sqrt{1-\eta_{\mathrm{rev}}}),1$, respectively, and $\alpha=1$ is fixed.}
\label{fig:EMP_etarev_alpha_delta}
\end{figure}
\begin{figure}[H]
        \centering		\includegraphics[width=0.48\textwidth]{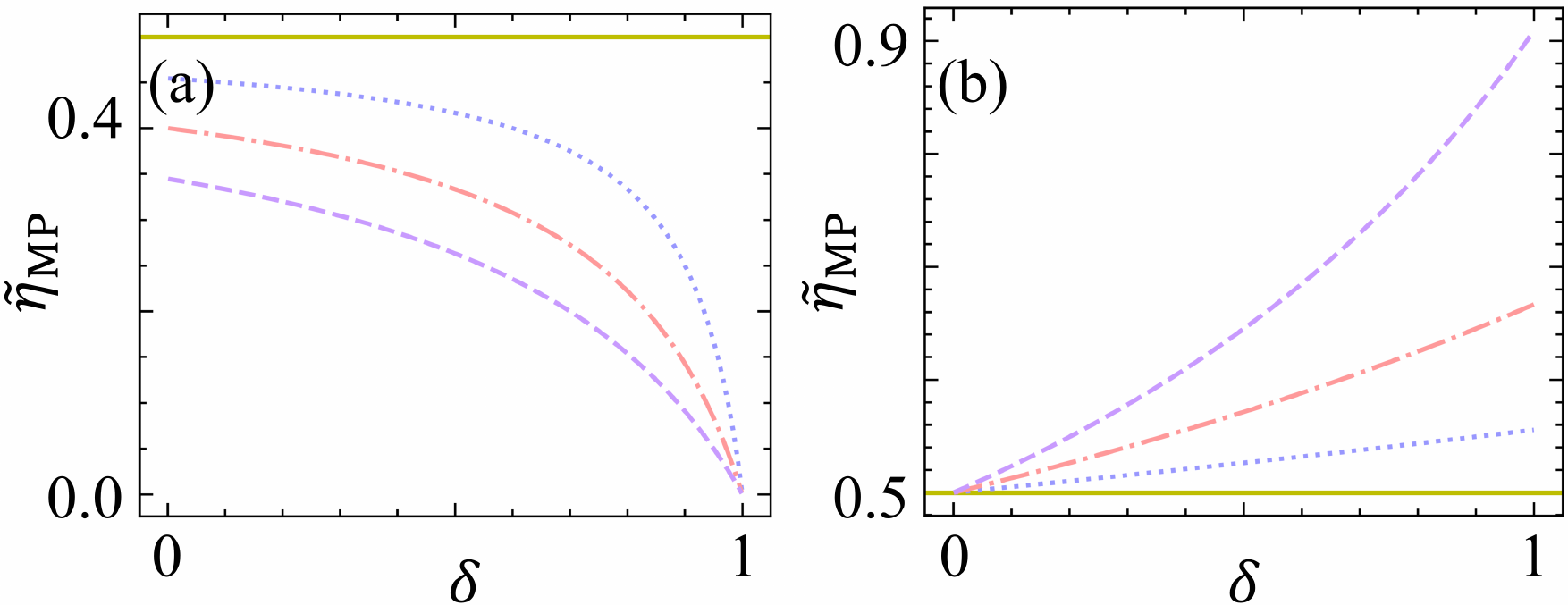}
	\caption{Normalized EMPs of refrigerators [(a)] and heat engines [(b)] as functions of $\delta$ in the case of $\alpha=1$. The blue dotted curve, red dash-dotted curve, and purple dashed curve are plotted with $\eta_{\rm{rev}}=0.2,0.5,0.9$, respectively. The green solid line represents $\tilde{\eta}=0.5$.}
\label{fig:EMP_delta_engine_refri}
\end{figure}

As the extreme point on the power-efficiency trade-off relation~\cite{maUniversalConstraintEfficiency2018}, EMP~\cite{tuEfficiencyMaximumPower2008,curzonEfficiencyCarnotEngine1975,vandenbroeckThermodynamicEfficiencyMaximum2005,espositoEfficiencyMaximumPower2010,zhao2025revisiting,lei2025universal} is a crucial parameter for evaluating finite-time heat engines ($\theta=\gamma=1$) and refrigerators ($\theta=\gamma=-1$). It follows from Eq.~\eqref{eq:tradeoff_general} that by solving $dP_{m}(\eta)/d\eta=0$, the corresponding EMP is
\begin{equation}
\eta_{\rm{MP}}=\frac{\alpha\eta_{\rm{rev}}}{1+\alpha-\chi \eta_{\rm{rev}}}.
    \label{eq:gene_EMP}
\end{equation}
Combining Eqs.~\eqref{eq:gene_EMP} and \eqref{chi}, the Normalized EMP $\Tilde{\eta}_{\rm{MP}}\equiv \eta_{\mathrm{MP}}/\eta_{\mathrm{rev}}$ of heat engines as a function of $\eta_{\mathrm{rev}}$ is plotted in Fig.~\ref{fig:EMP_etarev_alpha_delta} with different $\alpha$ [(a)] and $\delta$ [(b)]. It is shown in this figure that, for given $\eta_{\mathrm{rev}}$, increasing either $\alpha$ or $\delta$ can improve the EMP. Besides, it is worth mentioning that, due to the different signs (positive or negative) of $\chi$ corresponding to different thermal machines, the monotonicity of their EMP with respect to $\delta$ also varies, see Fig.~\ref{fig:EMP_delta_engine_refri} for a special case ($\alpha=1$) of refrigerators [(a)] and heat engines[(b)]. 

Furthermore, the leading-term coefficient of EMP with respect to $\eta_{\mathrm{rev}}$ is a noteworthy issue in the optimization of finite-time heat engines~\cite{vandenbroeckThermodynamicEfficiencyMaximum2005,tuEfficiencyMaximumPower2008}. In the near equilibrium regime, this coefficient is $1/2$~\cite{vandenbroeckThermodynamicEfficiencyMaximum2005,zhaiExperimentalTestPowerefficiency2023}, irrespective of engine details. However, such universality hinges on the validity of linear irreversible thermodynamics, which may be violated away from equilibrium~\cite{liang2023}. Here, as a result of Eq.~\eqref{eq:gene_EMP},
\begin{equation}
    \eta_{\rm{MP}}(\eta_{\rm{rev}}\ll1)\approx\frac{\alpha}{1+\alpha}\eta_{\rm{rev}},
\label{eq:gene_EMP1}
\end{equation}
which means that in our setup, the $1/2$-universality holds exclusively when $\alpha=1$. This fact is explicitly demonstrated in Fig.~\ref{fig:EMP_delta_engine_refri}(b): for heat engines with different $\delta$ under $\alpha=1$, their EMP converges exactly to  $\eta_{\rm{MP}}=\eta_{\mathrm{rev}}/2$ when  $\eta_{\mathrm{rev}}\to0$. In contrast, when $\alpha \to \infty$, EMP approaching $\eta_{\rm{rev}}$ follows directly from Eq.\eqref{eq:gene_EMP1}, and this trend is clearly shown in Fig.~\ref{fig:EMP_etarev_alpha_delta}(a).

\section{Specific trade-off relations for different \texorpdfstring{$\alpha$}{alpha}}
\label{IV}
As a specific application of the general results obtained above, we focus on two typical scenarios in this section: the slow-driving isothermal regime ($\alpha=1$) and the quantum finite-time adiabatic regime ($\alpha=2$), and derive the specific forms of trade-off relations for several thermal machines.

\subsection{The slowing driving isothermal regime of \texorpdfstring{$\alpha=1$}{alpha=1}}
\label{IVA}

In the slow-driving finite-time isothermal processes, the minimal energetic dissipation is proved to be $A_{\rm{min}}=\mathcal{L}^2\tau^{-1}$~\cite{brandnerThermodynamicGeometryMicroscopic2020}, where $\mathcal{L}$ is the thermodynamic length in thermodynamic geometry theory. This lower bound for irreversibility is achieved when the driving protocol of the cycle maintains uniform speed along the thermodynamic length trajectory (in the parametric space defined by thermodynamic state variables)~\cite{salamonThermodynamicLengthDissipated1983,brandnerThermodynamicGeometryMicroscopic2020,liGeodesicPathMinimal2022,chenOptimizingBrownianHeat2022}. With this $1/\tau$-scaling (i.e., $\alpha=1$), Eq.~\eqref{eq:tradeoff_general} becomes
\begin{equation}
P \le  \left(\frac{\rm{{I}_{\rm{rev}}}}{\mathcal{L}} \right)^2 \frac{\eta \left( \eta_{\rm{rev}}-\eta\right) \left( 1-\chi\eta_{\rm{rev}}\right)}{\theta\delta^{\frac{1+\theta}{2}}\chi^{-1}\left( 1-\chi\eta\right)^2},
\label{eq:tradeoff}
\end{equation}
the corresponding EMP $\eta^{\alpha=1}_{\rm{MP}}$ and maximum power $P^{\alpha=1}_{\rm{max}}$ are
\begin{equation}
    \eta^{\alpha=1}_{\rm{MP}}=\frac{\eta_{\rm{rev}}}{2-\chi \eta_{\rm{rev}}}
    \label{eq:MOF}
\end{equation}
and
 \begin{equation}
		P^{\alpha=1}_{\rm{max}}=\theta\gamma\left(\frac{\rm{I}_{\rm{rev}}\eta_{\rm{rev}}}{2\mathcal{L}}\right)^2\delta^{-\frac{1+\theta}{2}}\left| \chi  \right|,
		\label{eq:Pmax}
\end{equation}
respectively. Therefore, the normalized trade-off relation for a generic thermal machine operating in the slow-driving regime is
\begin{equation}
\Tilde{P} \le \frac{4\Tilde{\eta}(1-\Tilde{\eta})(1-\chi\eta_{\rm{rev}})}{(1-\chi\Tilde{\eta}\eta_{\rm{rev}})^2},
\label{eq:tradeoff_norm_alpha=1}
\end{equation}
where $\Tilde{P}\equiv P/P^{\alpha=1}_{\rm{max}}$ and $\Tilde{\eta}\equiv\eta/\eta_{\rm{rev}}$. We make two remarks here: 

i) the $1/\tau$-scaling of dissipation can also be derived using the framework of linear irreversible thermodynamics~\cite{wang2012efficiency,izumidaWorkOutputEfficiency2014,yuanOptimizingThermodynamicCycles2022}, hence our results are applicable to steady-state heat engines in linear response regime as well. Besides, utilizing the so-called shortcut to isothermality method~\cite{liShortcutsIsothermalityNonequilibrium2017,liGeodesicPathMinimal2022}, the $1/\tau$-scaling can be maintained even beyond the slow-driving regime~\cite{chenOptimizingBrownianHeat2022}, making our results applicable to any thermal machine constructed using shortcut to isothermality and not limited by driving speed.

ii) As a result of Eq.~\eqref{eq:Pmax}, the maximum power of thermal machines with $\theta\gamma<0$ ($\theta=-\gamma=-1$ for heat pumps and $\theta=-\gamma=1$ for heat exchangers) is negative, which is nonphysical. This implies that, under the condition of requiring positive output, the machines with $\theta\gamma<0$ do not have a locally optimal output power with respect to $\eta$, which results in a monotonic dependence of power on efficiency (trade-off relations with $\theta\gamma<0$ in Fig.~\ref{fig:alpha=1_tradeoff}). For thermal machines with $\theta\gamma>0$, the EMPs as functions of $\delta$ in the two-reservoirs case are plotted in Fig.~\ref{fig:EMP_delta_engine_refri}. In this figure, $\Tilde{\eta}_{\rm{MP}}=\eta_{\rm{MP}}/\eta_{\rm{rev}}\geq 0.5$ for heat engines while $\Tilde{\eta}_{\rm{MP}}\leq 0.5$ for refrigerators, different curve styles correspond to different $\eta_{\rm{rev}}$. In addition, Eq.~\eqref{eq:tradeoff_norm_alpha=1} implies that once $\Tilde{\eta}$ is fixed, the maximum  $\Tilde{P}$ is determined. In this scenario, increasing $P_{\rm{max}}$ can enhance $P$. As shown in Eq.~\eqref{eq:Pmax}, increasing $\rm{I}_{\rm{rev}}$ and decreasing $\mathcal{L}$ can effectively increase $P$, where the former can be achieved by enlarging the size of the machine (more working substance), and the latter can be realized through optimizing the driving path of the machine cycle~\cite{liGeodesicPathMinimal2022,chen2023geodesic} or leveraging collective advantages offered by interactions inside working substances~\cite{rolandiCollectiveAdvantagesFiniteTime2023,liang2023}. 

It follows from Eq.~\eqref{eq:tradeoff} that several typical trade-off relations for various thermal machines can be directly obtained, as demonstrated in Fig.~\ref{fig:alpha=1_tradeoff}. By specifying $\theta=1$ and $\gamma=1$, one has $\chi=\delta$ from Eq.~\eqref{chi}. Thus, it follows from Eq.~\eqref{eq:tradeoff_norm_alpha=1} that, the normalized trade-off relation for slow-driving heat engines reads
\begin{equation}
\Tilde{P} \le \frac{4 \Tilde{\eta}  \left( 1-\Tilde{\eta} \right)\left( 1-\delta \eta_{\rm{rev}} \right) }{\left( 1-\delta \eta_{\rm{rev}} \Tilde{\eta} \right)^2}
\label{eq:tradeoff_engine}
\end{equation}
with $\eta_{\rm{rev}} \equiv W_{\rm{rev}}/Q^{\rm{rev}}_{\rm{h}}$ the efficiency of a heat engine operating reversibly. This result is equivalent to that obtained with shortcut strategy~\cite{chenOptimizingBrownianHeat2022}. The equal sign of Eq.~\eqref{eq:tradeoff_engine} is saturated with $\delta=\mathcal{L}_{\rm{h}}/\mathcal{L}$ (See Appendix~\ref{appendixA} for an example of ideal gas), where $\mathcal{L}_{\rm{h}}$ is the total thermodynamic length contributed by all heat absorption processes~\cite{chenOptimizingBrownianHeat2022}. In addition, according to Eq.~\eqref{eq:MOF}, we obtain the EMP as
\begin{equation}
\eta_{\rm{MP}}=\frac{\eta_{\rm{rev}}}{2-\delta \eta_{\rm{rev}}},
\label{eq:EMP_engine}
\end{equation}
with $P_{\rm{max}}=\left( W_{\rm{rev}}\mathcal{L}^{-1}\right)^2/4$ the corresponding maximum power. Obviously from Eq.~\eqref{eq:EMP_engine} we have $\eta_{\rm{rev}}/2\le \eta_{\rm{MP}} \le \eta_{\rm{rev}}/(2-\eta_{\rm{rev}})$. This is consistent with the results obtained in Ref.~\cite{izumida2022irreversible,chenOptimizingBrownianHeat2022,van2013efficiency}.

When the scenario is further restricted to only two heat reservoirs, a high-temperature one with temperature $T_{\rm{h}}$ and a low-temperature one with temperature $T_{\rm{c}}$, the reversible efficiency simplifies to its Carnot form, namely, $\eta_{\rm{rev}}=\eta_{\rm{C}}=1-T_{\rm{c}}/T_{\rm{h}}$. In this case, Eq.~\eqref{eq:EMP_engine} is simplified as $\eta_{\rm{C}}/(2-\delta\eta_{\rm{C}})$, recovering the EMP of low-dissipation Carnot engine obtained by Esposito et al.~\cite{espositoEfficiencyMaximumPower2010}. The EMP in this case has been plotted in Fig.~\ref{fig:EMP_etarev_alpha_delta}(b), where the blue dotted and purple dashed curves represent the upper bound $\eta_{\rm{C}}/(2-\eta_{\rm{C}})$ ($\delta=1$) and lower bound $\eta_{\rm{C}}/2$ ($\delta=0$), respectively. The red dash-dotted is associated with $\delta=1/(1+\sqrt{1-\eta_{\mathrm{C}}})$, representing the Curzon-Ahlborn efficiency~\cite{curzonEfficiencyCarnotEngine1975,zhao2025revisiting}

Beside, by taking $\delta=1$, the overall upper bound for efficiency at arbitrary given power is obtained from Eq.~\eqref{eq:tradeoff_engine} as (See Appendix~\ref{appendixB} for details)
\begin{equation}
\Tilde{\eta}_{\rm{U}}=1-\frac{\left(1-\eta_{\mathrm{C}}\right) \Tilde{P}}{2(1+\sqrt{1-\Tilde{P}})-\Tilde{P} \eta_{\mathrm{C}}},
\label{eq:eta_Ua}  
\end{equation}
and the overall lower bound, $\Tilde{\eta}_{\rm{L}}=(1-\sqrt{1-\Tilde{P}})/2$, is achieved with $\delta=0$. These bounds have been obtained before with different approaches~\cite{holubecEfficiencyMaximumPower2015,holubecMaximumEfficiencyLowdissipation2016,maUniversalConstraintEfficiency2018} and experimentally tested~\cite{zhaiExperimentalTestPowerefficiency2023}. Furthermore, in the case of two finite-sized heat reservoirs, the quasi-static efficiency of the heat engine is referred to as efficiency at maximum work ($\eta_{\rm{MW}}$)~\cite{ondrechenMaximumWorkFinite1981,maEffectFiniteSizeHeat2020}. In this scenario, it can be easily verified that Eq.~\eqref{eq:tradeoff_engine} is consistent with the result obtained with linear irreversible thermodynamics, as given by [Eq. (10)] of Ref.~\cite{yuanOptimizingThermodynamicCycles2022}. The EMP of the engine satisfies $\eta_{\rm{MW}}/2\le \eta_{\rm{MP}} \le \eta_{\rm{MW}}/(2-\eta_{\rm{MW}})$~\cite{amelkinMaximumPowerProcesses2004,izumidaWorkOutputEfficiency2014,yuanOptimizingThermodynamicCycles2022}. We do not delve into detailed discussions on this issue here.

For refrigerators, their efficiency is referred to as the COP, denoted by the symbol $\varepsilon$. It follows from Eq.~\eqref{eq:tradeoff_norm_alpha=1} that, by specifying $\theta=-1$ and $\gamma=-1$, i.e., $\chi=-(1-\delta)^{-1}$, and using variable substitutions $\eta \rightarrow \varepsilon$ and $\eta_{\rm{rev}}\rightarrow \varepsilon_{\rm{rev}}$, the trade-off relation for slow-driving refrigerators is obtained as
\begin{equation}
\Tilde{P} \le\frac{ 4 \Tilde{\varepsilon}  \left( 1-\Tilde{\varepsilon} \right)\left[ 1+(1-\delta)^{-1} \varepsilon_{\rm{rev}} \right] }{\left[1+(1-\delta)^{-1} \varepsilon_{\rm{rev}} \Tilde{\varepsilon} \right]^2},
\label{eq:tradeoff_refrigerator}
\end{equation}
where $\varepsilon_{\rm{rev}} \equiv Q_{\rm{rev}}^{\rm{c}}/W_{\rm{rev}}$ represents the COP a refrigerator operating reversibly. In this situation, the efficiency at maximum cooling power and the corresponding maximum cooling power follow
\begin{equation}
\varepsilon_{\rm{MP}}=\frac{\varepsilon_{\rm{rev}}}{2+(1-\delta)^{-1}\varepsilon_{\rm{rev}}},P_{\rm{max}}=\frac{(1-\delta)^{-1}}{4}\left( \frac{W_{\rm{rev}}\varepsilon_{\rm{rev}}}{\mathcal{L}} \right)^2,
\label{eq:EMP_refrigerator}
\end{equation}
respectively. Since $(1-\delta)^{-1}\in[1,\infty)$, the EMP of refrigerators satisfies $0\le \varepsilon_{\rm{MP}} \le \varepsilon_{\rm{rev}}/(2+\varepsilon_{\rm{rev}})$. For given reversible cycle qualities, the maximum cooling power and its corresponding COP are solely determined by the dissipation ratio $\delta$. We note that as the dissipation at the high-temperature end diminishes ($\delta\rightarrow1$), the cooling power diverges, whereas the COP approaches zero at this point. However, the product of the two is a constant, namely, $\lim_{\delta\rightarrow1}\left(P_{\rm{max}}\epsilon_{\rm{MP}}\right)=[W_{\rm{rev}}\varepsilon_{\rm{rev}}/(2\mathcal{L})]^2$, indicating that when the dissipation at the high-temperature heat reservoir is extremely small, the optimized extreme point moves along a hyperbola. In more specific case with two heat baths, one has $\eta_{\rm{rev}}=\epsilon_{\rm{C}}=T_{\rm{c}}/\left( T_{\rm{h}}-T_{\rm{c}} \right)$ and the corresponding EMP is $\epsilon_{\rm{MP}}=\epsilon_{\rm{C}}/[2+(1-\delta)^{-1} \epsilon_{\rm{C}}]$, 
\label{eq:EMP_LDrefrigerator}
which is consisting with Ref.~\cite{holubecMaximumEfficiencyLowdissipation2020}. In Fig.~\ref{fig:EMP_delta_engine_refri}(a), $\epsilon_{\rm{MP}}$ as a function of $\delta$ with $\eta_{\rm{C}}=0.2,0.5,0.9$ is plotted as the curves below $\Tilde{\eta}_{\rm{MP}}\leq 0.5$, larger $\epsilon_{\rm{C}}$ leads to lower $\tilde{\epsilon}_{\rm{MP}}$.

From Eq.~\eqref{eq:tradeoff}, the trade-off relation of heat pump is (see Fig.~\ref{fig:alpha=1_tradeoff})
\begin{equation}
P \le  \left(\frac{\rm{{I}_{\rm{rev}}}}{\mathcal{L}} \right)^2 \frac{\eta \left( \eta_{\rm{rev}}-\eta\right) \left(\eta_{\rm{rev}}-\delta\right)}{\left(\eta-\delta\right)^2},
\label{eq:tradeoff_pump}
\end{equation}
As we mentioned in remark ii) below Eq.~\eqref{eq:tradeoff_norm_alpha=1}, heat pumps and heat exchangers with $\theta\gamma<0$ do not have locally maximum positive output power. The monotonicity of their trade-off relations can be directly observed from Fig.~\ref{fig:alpha=1_tradeoff}. Mathematically, $\theta\gamma<0$ leads to $P$ being a monotonically increasing function of cycle time $\tau$ in the physically meaningful region of $\tau>0$, without any optimal time that locally maximizes $P$. This is also revealed in the specific analysis of heat pumps in Ref.~\cite{yeMaximumEfficiencyLowdissipation2022}.

\subsection{The finite-time quantum adiabatic regime of \texorpdfstring{$\alpha=2$}{alpha=2}}
\label{IVB}
It has been proved that $\alpha=2$ corresponds to the finite-time adiabatic processes where $A_{\rm{min}}=\Sigma/\tau^2$~\cite{chen2019achieve}. In this case, Eq.~\eqref{eq:tradeoff_general} becomes
\begin{equation}
P\le  \frac{\rm{{I}^{3/2}_{\rm{rev}}}\eta\left( 1-\chi\eta_{\rm{rev}}\right)}{\sqrt{\Sigma\theta\delta^{1+\theta}}} \sqrt{\frac{\chi\left(\eta_{\rm{rev}}-\eta\right)}{\left( 1-\chi\eta\right)^{^{3}}}}.
\label{eq:A_L_P_1_adia}
\end{equation}
This serves as the trade-off relation for finite-time quantum Otto thermal machines, which has not been obtained before, to the best of our knowledge.
\begin{figure}[h]
        \centering		\includegraphics[width=0.48\textwidth]{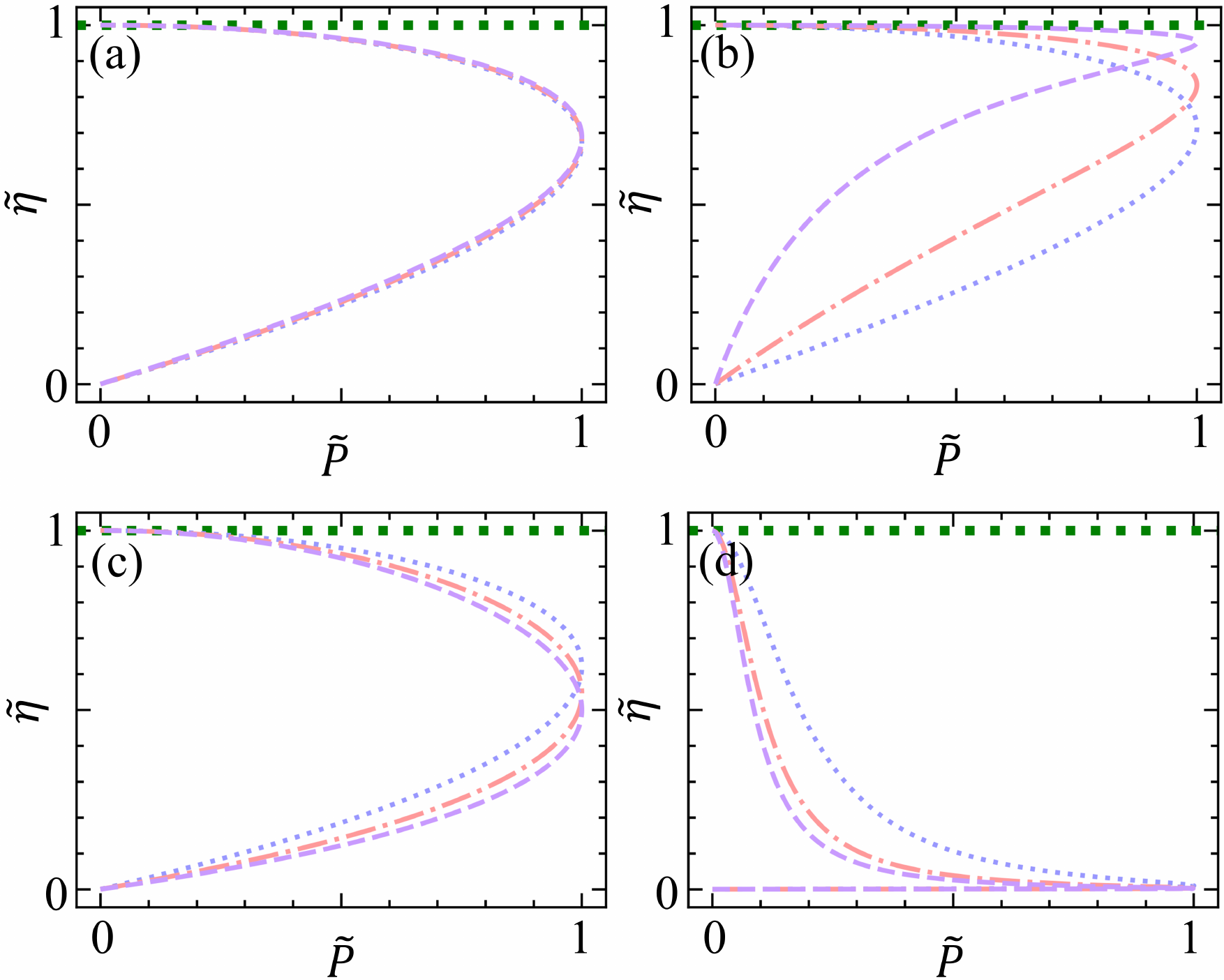}
	\caption{The normalized trade-off relations associated with $\alpha=2$ for heat engines[(a-b)] and refrigerators[(c-d)]. The blue dotted, red dash-dotted, and purple dashed curve are plotted with $\eta_{\mathrm{rev}}=0.2,0.6,0.9$, respectively. $\Tilde{\eta}=1$ is plotted with the green dotted line. $\delta=0.1$ is used to plot (a) and (c), while $\delta=1$ for (b) and (d).}
\label{fig:etar_engine_refri}
\end{figure}

Figure~\ref{fig:etar_engine_refri} illustrates trade-off relations with different $\eta_{\mathrm{rev}}$ and $\delta$. As can be seen from Fig.~\ref{fig:etar_engine_refri}(a-b), the differences in trade-off relations of heat engines caused by different $\eta_{\mathrm{rev}}$ increase with the increase of $\delta$. Meanwhile, a larger $\delta$ leads to a higher EMP of heat engines. In contrast, increasing $\delta$ shifts the trade-off relations of refrigerators downward and reduces their EMPs, as shown in Fig.~\ref{fig:etar_engine_refri}(c-d). This implies that heat engines and refrigerators require different optimization strategies with respect to $\delta$. In addition, substituting $\alpha=2$ into Eq.~\eqref{eq:gene_EMP} yields
\begin{equation}    \eta_{\rm{MP}}=\frac{2\eta_{\rm{rev}}}{3-\chi \eta_{\rm{rev}}},
    \label{eq:MOF_alpha=2}
\end{equation}
which is consistent with the result of finite-time quantum Otto heat engines ($\chi=\delta$) given in Ref.~\cite{chen2019achieve}. By comparing this result with Eq.~\eqref{eq:EMP_engine}, the fact that the quantum Otto engine can exceed the Carnot cycle’s EMP with the same $\delta$ indicates that out-of-equilibrium quantum thermodynamic cycles may have greater advantages in finite-time optimization under specific conditions~\cite{chen2019achieve,chenBoostingPerformanceQuantum2019,fei2022efficiency}.

\section{Conclusion}
\label{V}
In this work, we have proposed a unified framework for describing arbitrary thermal machines, addressing a key limitation in previous studies: the requirement for separate performance parameter definitions for machines with distinct functions, which caused practical inconvenience. The core of our approach lies in two key points: i) unifying performance descriptions across different thermal machines using sign functions, and ii) establishing relations among total cycle dissipation, cycle time, and the machine's power and efficiency. This unified framework is theoretically applicable to any form of finite-time machine dissipation $A(\tau)$. 

The typical dissipation $A(\tau)\propto\tau^{-\alpha}$ draws critical benefits from control strategies, namely optimizing the regulation scheme of the system-environment interaction~\cite{pancotti2020speed}. Leveraging the unified thermal machines description, we derived a general power-efficiency trade-off relation [Eq.\eqref{eq:tradeoff_general}] for generic thermal machines exhibiting $\tau^{-\alpha}$-type cycle irreversibility. We further demonstrate the controllability of this trade-off relation in different cases with specified parameters $\alpha$, $\theta$, and $\gamma$. Our general result recovers several previous results [Sec.~\ref{IV}], yields the complete performance curve, and enables the investigation of a broader range of operating regimes for thermal machines. Notably, when $\alpha$ approaches infinity, theoretical derivations indicate that thermal machines can simultaneously attain reversible efficiency and maximum power[Sec.~\ref{IIIB}]. This outcome not only validates the universal analytical capability of this method for thermal machine behavior across different scaling regimes but also makes innovative predictions regarding the thermodynamic characteristics under extreme scaling limits. We have also determined the trade-off relation of finite-time quantum adiabatic Otto engines[Sec.~\ref{IVB}] for the first time. This study lays the foundation for a unified description of thermodynamic cycles and general approaches to thermal machine performance optimization.

As a final remark, the right-angled triangle trade-off relation shown in Fig.~\ref{fig:tradeoff_alpha_engine}(a) is similar to the optimal trade-off relation for the degenerate heat engine~\cite{liang2023}. This suggests that regulating the thermal machine-reservoir coupling (current work and Ref.~\cite{pancotti2020speed}) and regulating the intrinsic structure of the machine's working substance (Ref.~\cite{liang2023}) may have a more profound connection in the context of optimizing machine performance.

\section*{Acknowledgment}
This work is supported by the National Natural Science Foundation of China under grant No. 12305037, and the Fundamental Research Funds for the Central Universities under grant No. 2023NTST017.

\appendix
\section{Thermodynamic geometry of ideal gas}
\label{appendixA}

In the following, we specifically demonstrate how to use the thermodynamic geometric approach to determine the dissipation ratio $\delta$ in a finite-time Carnot cycle with an ideal gas as the working substance. When the gas with a molar amount of $n$ and volume $V$ is driven slowly in a hot reservoir of temperature $T_{\rm{h}}$ (finite-time high-temperature isothermal process of duration $\tau_{\rm{h}}$), the gas pressure $p$ reads~\cite{maExperimentalTestTau2020}
\begin{equation}
p=p_{\rm{q}}-\frac{T_{\rm{h}}}{\kappa_{\rm{h}}} \left(\frac{nR}{V}\right)^2\dot{V},
\label{eq:slowdriving_equation} 
\end{equation}
where $p_{\rm{q}}$ is the gas pressure in quasi-static process($\dot{V}\rightarrow0$), $\kappa_{\rm{h}}$ is the heat transfer coefficient between the gas and the reservoir, and $R$ is the ideal gas constant. According to Ref.~\cite{chen2021extrapolating},
\begin{align*}
\mathcal{L}_{\rm{h}} & =\int_{0}^{\tau_{1}} \sqrt{\dot W_{\rm{ex}} (\textit{t})} \rm{d} \textit{t}=\int_{0}^{\tau_{1}} \sqrt{-\textit{p}_{\rm{ex}} (\textit{t}) \dot{\textit{V}}} \rm{d} \textit{t} \\
& =\left|\int_{V_{0}}^{V_{f}} \sqrt{\frac{T_{\rm{h}}}{\kappa_{\rm{h}}}} \frac{n R d V}{V}\right|=n R \sqrt{\frac{T_{\rm{h}}}{\kappa_{\rm{h}}}}\left|\ln \left(\frac{V_{f}}{V_{0}}\right)\right|,
\end{align*}
where $p_{\rm{ex}}\equiv p-p_{\rm{q}}$ has been used. Thus, the corresponding dissipation in this process follows as~\cite{salamonThermodynamicLengthDissipated1983} 
\begin{equation}
Q_{\rm{ir}}^{\rm{h}} \ge \frac{\mathcal{L}_{\rm{h}}^2}{\tau_{\rm{h}}}=\frac{T_{\rm{h}}\left[n R \ln \left(V_{f} / V_{0}\right)\right]^{2}}{\kappa_{\rm{h}} \tau_{\rm{h}}}.
\label{eq:High_T_diss}
\end{equation}
Similarly, in the low-temperature finite-time isothermal process of duration $\tau_{\rm{c}}$
\begin{equation}
Q_{\rm{ir}}^{\rm{c}} \ge \frac{\mathcal{L}_{\rm{c}}^2}{\tau_{\rm{c}}}=\frac{T_{\rm{c}}\left[n R \ln \left(V_{f} / V_{0}\right)\right]^{2}}{\kappa_{\rm{c}} \tau_{\rm{c}}}.
\label{eq:Low_T_diss}
\end{equation}
For the whole cycle, the minimal dissipation is~\cite{brandnerThermodynamicGeometryMicroscopic2020,chenOptimizingBrownianHeat2022}
\begin{equation}
A_{\rm{min}}(\tau=\tau_{\rm{h}}+\tau_{\rm{c}})=\frac{\left(\mathcal{L}_{\rm{h}}+\mathcal{L}_{\rm{c}}\right)^2}{\tau_{\rm{h}}+\tau_{\rm{c}}}.
\end{equation}
It is easily checked that
\begin{align*}
& \frac{\mathcal{L}_{\rm{h}}^2}{\tau_{\rm{h}}}+\frac{\mathcal{L}_{\rm{C}}^2}{\tau_{\rm{C}}}-\frac{\left(\mathcal{L}_{\rm{h}}+\mathcal{L}_{\rm{c}}\right)^2}{\tau_{\rm{h}}+\tau_{\rm{c}}} \\
& =\frac{\mathcal{L}_{\rm{h}}^2 \tau_{\rm{c}}\left(\tau_{\rm{h}}+\tau_{\rm{c}}\right)+\mathcal{L}_{\rm{c}}^2 \tau_{\rm{h}}\left(\tau_{\rm{h}}+\tau_{\rm{c}}\right)-\left(\mathcal{L}_{\rm{h}}+\mathcal{L}_{\rm{c}}\right)^2 \tau_{\rm{h}} \tau_{\rm{c}}}{\tau_{\rm{h}}\tau_{\rm{c}}\left(\tau_{\rm{h}}+\tau_{\rm{c}}\right)} \\
& =\frac{\left(\mathcal{L}_{\rm{h}} \tau_{\rm{c}}-\mathcal{L}_{\rm{c}} \tau_{\rm{h}}\right)^2}{\tau_{\rm{h}} \tau_{\rm{c}}\left(\tau_{\rm{h}}+\tau_{\rm{c}}\right)} \geq 0,
\end{align*}
namely, the total dissipation $A=A_{\rm{h}}+A_{\rm{c}} \geq A_{\rm{min}}$. Here, the equal sign is taken when
\begin{equation}
A_{\rm{h}}=\frac{\mathcal{L}_{\rm{h}}^2}{\tau_{\rm{h}}},~~A_{\rm{c}}=\frac{\mathcal{L}_{\rm{c}}^2}{\tau_{\rm{c}}},~~\frac{\mathcal{L}_{\rm{h}}}{\mathcal{L}_{\rm{c}}}=\frac{\tau_{\rm{h}}}{\tau_{\rm{c}}},
\label{eq:total_diss_equa}
\end{equation}
which is specifically written, for ideal gas, as
\begin{equation}
\frac{A_{\rm{h}}}{A_{\rm{c}}}=\frac{\mathcal{L}_{\rm{h}}}{\mathcal{L}_{\rm{c}}}=\frac{\tau_{\rm{h}}}{\tau_{\rm{c}}}=\sqrt{\frac{T_{\rm{h}}\kappa_{\rm{c}}}{T_{\rm{c}}\kappa_{\rm{h}}}}.
\label{eq:total_diss_equa_3}
\end{equation}
Therefore, we obtain
\begin{equation}
		\delta=\frac{\mathcal{L}_{\rm{h}}}{\mathcal{L}}=\frac{1}{1+\sqrt{\frac{T_{\rm{c}}\kappa_{\rm{h}}}{T_{\rm{h}}\kappa_{\rm{c}}}}}.
		\label{eq:CA}
	\end{equation}
Since $\kappa_{\rm{h}}/\kappa_{\rm{c}}\in(0,\infty)$, one has $\delta\in(0,1)$. Especially, in the  symmetric case with $\kappa_{\rm{h}}/\kappa_{\rm{c}}=1$, it follows from the EMP equation of $\alpha=1$ that $\eta_{\rm{MP}}=1-\sqrt{1-\eta_{\rm{C}}}$, which is exactly the Curzon-Ahlborn efficiency~\cite{curzonEfficiencyCarnotEngine1975}.

\section{Upper bound of efficiency from Eq.~\eqref{eq:tradeoff_engine}} 
\label{appendixB}
The range of EMP indicates a varying trade-off relation in heat engines as a function of the dissipation ratio, allowing us to establish tighter overall upper and lower bounds of efficiency at given power from Eq.~\eqref{eq:tradeoff_engine}
\begin{equation}
\frac{\Tilde{\eta}  \left( 1-\Tilde{\eta} \right)}{\left( 1-\delta \eta_{\rm{rev}} \Tilde{\eta} \right)^2} \geq \frac{\Tilde{P} }{4\left( 1-\delta \eta_{\rm{rev}} \right)}\equiv \lambda,
\label{eq:eta}
\end{equation}
namely,
\begin{equation}
\left( 1+\lambda \delta^2 \eta_{\rm{rev}}^2 \right)\Tilde{\eta}^2 -\left( 1+2\lambda \delta \eta_{\rm{rev}} \right) \Tilde{\eta}+\lambda \le 0.
\label{eq:eta_eq}
\end{equation}
The solution of the above inequality of $\Tilde{\eta}$ are
\begin{equation}
\Tilde{\eta}_{\pm}=\frac{\left(1+2 \lambda \delta \eta_{\rm{rev}}\right) \pm \sqrt{4 \lambda \delta \eta_{\rm{rev}}+1-4 \lambda}}{2\left(\lambda \delta^{2} \eta_{\rm{rev}}^{2}+1\right)}.
\label{eq:eta_root}
\end{equation}
Under the condition that the power and thermodynamic parameters are fixed, $\Tilde{\eta}_{\pm}$ delineate the upper and lower bounds of the low-dissipation region. Generally, precise knowledge of the heat engine parameters is elusive; hence, deriving universal upper/lower bounds that are valid for all $\delta$ within this range is of practical significance. Next, we prove the monotonic increasing nature of the tight upper and lower bounds for efficiency, as a function of a given power with a variable parameter $\delta$.

To investigate the monotonicity of this function, we differentiate it with respect to $\delta$, yielding the following equation
\begin{equation}
\Tilde{\eta}_{-}=\frac{\left(1+2 \lambda \delta \eta_{\rm{rev}}\right) - \sqrt{4 \lambda \delta \eta_{\rm{rev}}+1-4 \lambda}}{2\left(\lambda \delta^{2} \eta_{\rm{rev}}^{2}+1\right)},
\label{eq:deta-_}
\end{equation}
such that

\begin{equation}
\frac{\rm{d}\Tilde{\eta}_{-}}{\rm{d} \delta} = \frac{-\eta_{\rm{rev}} \Tilde{P} \left( a_{-} x^2 + b_{-} x -4\right)}{\left( \Tilde{P}x^2-4 x+4 \right)^2},
\label{eq:deta-_alpha}
\end{equation}
where $a_{-}=\Tilde{P}+ 2\sqrt{1-\Tilde{P}} -2,~b_{-}= 4- 4\sqrt{1-\Tilde{P}},~x=\delta \eta_{\rm{rev}}$. The sign of the above equation is determined by $a_{-} x^2 + b_{-} x -4$. Notably, it is straightforward to observe that $a_{-} \le 0$, with equality holding only when $\Tilde{P}=1$. Furthermore, under the condition $\Delta=b_{-}^2+16a_{-}=0$, we find that $a_{-} x^2 + b_{-} x -4 \le 0$ for $\Tilde{P} \ne 1$. When $\Tilde{P}=1$ both $a_{-}$ and $b_{-}$ vanish, reducing the quadratic expression as $-4$. We conclude that $\frac{\rm{d}\Tilde{\eta}_{-}}{\rm{d} \delta} > 0$ for $\delta \in [0,1]$. Thus, the overall lower bound is reached at $\delta=0$, substituting which into Eq.~\eqref{eq:deta-_alpha} yields $\Tilde{\eta}_{\rm{L}}=(1-\sqrt{1-\Tilde{P}})/2$. Similarly, for
\begin{equation}
\Tilde{\eta}_{+}=\frac{\left(1+2 \lambda \delta \eta_{\rm{rev}}\right) + \sqrt{4 \lambda \delta \eta_{\rm{rev}}+1-4 \lambda}}{2\left(\lambda \delta^{2} \eta_{\rm{rev}}^{2}+1\right)},
\label{eq:eta+_alpha}
\end{equation}
one has
\begin{equation}
\frac{\rm{d}\Tilde{\eta}_{+}}{\rm{d} \delta} = \frac{-\eta_{\rm{rev}} \Tilde{P} \left( a_{+} x^2 + b_{+} x -4\right)}{\left( \Tilde{P}x^2-4 x+4 \right)^2},
\label{eq:deta+_alpha}
\end{equation}
where $a_{+}=\Tilde{P}- 2\sqrt{1-\Tilde{P}} -2,~b_{+}= 4+ 4\sqrt{1-\Tilde{P}},~x=\delta \eta_{\rm{rev}}$. The sign of the above equation is determined by $a_{+} x^2 + b_{+} x -4$. Notably, it is straightforward to observe that $a_{+} < 0$. Furthermore, under the condition $\Delta=b_{+}^2+16a_{+}=0$, we find that $a_{+} x^2 + b_{+} x -4 \le 0$. In summary, we conclude that $\frac{\rm{d}\Tilde{\eta}_{+}}{\rm{d} \delta} \geq 0$ for $\delta \in [0,1]$. Therefore, the overall upper bound $\Tilde{\eta}_{\rm{U}} \ge \Tilde{\eta}$ is achieved with $\delta=1$, substituting which into Eq.~\eqref{eq:eta+_alpha}, Eq.~\eqref{eq:eta_Ua} is obtained. 

\bibliography{main}

\begin{thebibliography}{62}%
\makeatletter
\providecommand \@ifxundefined [1]{%
 \@ifx{#1\undefined}
}%
\providecommand \@ifnum [1]{%
 \ifnum #1\expandafter \@firstoftwo
 \else \expandafter \@secondoftwo
 \fi
}%
\providecommand \@ifx [1]{%
 \ifx #1\expandafter \@firstoftwo
 \else \expandafter \@secondoftwo
 \fi
}%
\providecommand \natexlab [1]{#1}%
\providecommand \enquote  [1]{``#1''}%
\providecommand \bibnamefont  [1]{#1}%
\providecommand \bibfnamefont [1]{#1}%
\providecommand \citenamefont [1]{#1}%
\providecommand \href@noop [0]{\@secondoftwo}%
\providecommand \href [0]{\begingroup \@sanitize@url \@href}%
\providecommand \@href[1]{\@@startlink{#1}\@@href}%
\providecommand \@@href[1]{\endgroup#1\@@endlink}%
\providecommand \@sanitize@url [0]{\catcode `\\12\catcode `\$12\catcode `\&12\catcode `\#12\catcode `\^12\catcode `\_12\catcode `\%12\relax}%
\providecommand \@@startlink[1]{}%
\providecommand \@@endlink[0]{}%
\providecommand \url  [0]{\begingroup\@sanitize@url \@url }%
\providecommand \@url [1]{\endgroup\@href {#1}{\urlprefix }}%
\providecommand \urlprefix  [0]{URL }%
\providecommand \Eprint [0]{\href }%
\providecommand \doibase [0]{https://doi.org/}%
\providecommand \selectlanguage [0]{\@gobble}%
\providecommand \bibinfo  [0]{\@secondoftwo}%
\providecommand \bibfield  [0]{\@secondoftwo}%
\providecommand \translation [1]{[#1]}%
\providecommand \BibitemOpen [0]{}%
\providecommand \bibitemStop [0]{}%
\providecommand \bibitemNoStop [0]{.\EOS\space}%
\providecommand \EOS [0]{\spacefactor3000\relax}%
\providecommand \BibitemShut  [1]{\csname bibitem#1\endcsname}%
\let\auto@bib@innerbib\@empty
\bibitem [{\citenamefont {Andresen}\ \emph {et~al.}(1977)\citenamefont {Andresen}, \citenamefont {Berry}, \citenamefont {Nitzan},\ and\ \citenamefont {Salamon}}]{andresenThermodynamicsFiniteTime1977a}%
  \BibitemOpen
  \bibfield  {author} {\bibinfo {author} {\bibfnamefont {B.}~\bibnamefont {Andresen}}, \bibinfo {author} {\bibfnamefont {R.~S.}\ \bibnamefont {Berry}}, \bibinfo {author} {\bibfnamefont {A.}~\bibnamefont {Nitzan}},\ and\ \bibinfo {author} {\bibfnamefont {P.}~\bibnamefont {Salamon}},\ }\bibfield  {title} {\bibinfo {title} {Thermodynamics in finite time. {{I}}. {{The}} step-{{Carnot}} cycle},\ }\href {https://doi.org/10.1103/PhysRevA.15.2086} {\bibfield  {journal} {\bibinfo  {journal} {Phys. Rev. A}\ }\textbf {\bibinfo {volume} {15}},\ \bibinfo {pages} {2086} (\bibinfo {year} {1977})}\BibitemShut {NoStop}%
\bibitem [{\citenamefont {Salamon}\ \emph {et~al.}(1977)\citenamefont {Salamon}, \citenamefont {Andresen},\ and\ \citenamefont {Berry}}]{salamonThermodynamicsFiniteTime1977}%
  \BibitemOpen
  \bibfield  {author} {\bibinfo {author} {\bibfnamefont {P.}~\bibnamefont {Salamon}}, \bibinfo {author} {\bibfnamefont {B.}~\bibnamefont {Andresen}},\ and\ \bibinfo {author} {\bibfnamefont {R.~S.}\ \bibnamefont {Berry}},\ }\bibfield  {title} {\bibinfo {title} {Thermodynamics in finite time. {{II}}. {{Potentials}} for finite-time processes},\ }\href {https://doi.org/10.1103/PhysRevA.15.2094} {\bibfield  {journal} {\bibinfo  {journal} {Phys. Rev. A}\ }\textbf {\bibinfo {volume} {15}},\ \bibinfo {pages} {2094} (\bibinfo {year} {1977})}\BibitemShut {NoStop}%
\bibitem [{\citenamefont {Ondrechen}\ \emph {et~al.}(1980)\citenamefont {Ondrechen}, \citenamefont {Berry},\ and\ \citenamefont {Andresen}}]{ondrechenThermodynamicsFiniteTime1980}%
  \BibitemOpen
  \bibfield  {author} {\bibinfo {author} {\bibfnamefont {M.~J.}\ \bibnamefont {Ondrechen}}, \bibinfo {author} {\bibfnamefont {R.~S.}\ \bibnamefont {Berry}},\ and\ \bibinfo {author} {\bibfnamefont {B.}~\bibnamefont {Andresen}},\ }\bibfield  {title} {\bibinfo {title} {Thermodynamics in finite time: {{A}} chemically driven engine},\ }\href {https://doi.org/10.1063/1.439744} {\bibfield  {journal} {\bibinfo  {journal} {J. Chem. Phys.}\ }\textbf {\bibinfo {volume} {72}},\ \bibinfo {pages} {5118} (\bibinfo {year} {1980})}\BibitemShut {NoStop}%
\bibitem [{\citenamefont {Qiu}\ \emph {et~al.}(2025)\citenamefont {Qiu}, \citenamefont {Nomura}, \citenamefont {Zhang}, \citenamefont {Lu}, \citenamefont {Volz}, \citenamefont {Chen}, \citenamefont {Zhang}, \citenamefont {Zhang}, \citenamefont {Woods}, \citenamefont {Dai} \emph {et~al.}}]{qiu2025roadmap}%
  \BibitemOpen
  \bibfield  {author} {\bibinfo {author} {\bibfnamefont {Y.}~\bibnamefont {Qiu}}, \bibinfo {author} {\bibfnamefont {M.}~\bibnamefont {Nomura}}, \bibinfo {author} {\bibfnamefont {Z.}~\bibnamefont {Zhang}}, \bibinfo {author} {\bibfnamefont {S.}~\bibnamefont {Lu}}, \bibinfo {author} {\bibfnamefont {S.}~\bibnamefont {Volz}}, \bibinfo {author} {\bibfnamefont {J.}~\bibnamefont {Chen}}, \bibinfo {author} {\bibfnamefont {J.}~\bibnamefont {Zhang}}, \bibinfo {author} {\bibfnamefont {H.}~\bibnamefont {Zhang}}, \bibinfo {author} {\bibfnamefont {L.~M.}\ \bibnamefont {Woods}}, \bibinfo {author} {\bibfnamefont {G.}~\bibnamefont {Dai}}, \emph {et~al.},\ }\bibfield  {title} {\bibinfo {title} {Roadmap on thermodynamics and thermal metamaterials},\ }\href {https://doi.org/10.15302/frontphys.2025.065500} {\bibfield  {journal} {\bibinfo  {journal} {Front. Phys.}\ }\textbf {\bibinfo {volume} {20}},\ \bibinfo {pages} {065500} (\bibinfo {year} {2025})}\BibitemShut {NoStop}%
\bibitem [{\citenamefont {Chen}\ and\ \citenamefont {Yan}(1989)}]{chenEffectHeatTransfer1989}%
  \BibitemOpen
  \bibfield  {author} {\bibinfo {author} {\bibfnamefont {L.}~\bibnamefont {Chen}}\ and\ \bibinfo {author} {\bibfnamefont {Z.}~\bibnamefont {Yan}},\ }\bibfield  {title} {\bibinfo {title} {The effect of heat-transfer law on performance of a two-heat-source endoreversible cycle},\ }\href {https://doi.org/10.1063/1.455832} {\bibfield  {journal} {\bibinfo  {journal} {J. Chem. Phys.}\ }\textbf {\bibinfo {volume} {90}},\ \bibinfo {pages} {3740} (\bibinfo {year} {1989})}\BibitemShut {NoStop}%
\bibitem [{\citenamefont {Holubec}\ and\ \citenamefont {Ryabov}(2015)}]{holubecEfficiencyMaximumPower2015}%
  \BibitemOpen
  \bibfield  {author} {\bibinfo {author} {\bibfnamefont {V.}~\bibnamefont {Holubec}}\ and\ \bibinfo {author} {\bibfnamefont {A.}~\bibnamefont {Ryabov}},\ }\bibfield  {title} {\bibinfo {title} {Efficiency at and near maximum power of low-dissipation heat engines},\ }\href {https://doi.org/10.1103/PhysRevE.92.052125} {\bibfield  {journal} {\bibinfo  {journal} {Phys. Rev. E}\ }\textbf {\bibinfo {volume} {92}},\ \bibinfo {pages} {052125} (\bibinfo {year} {2015})}\BibitemShut {NoStop}%
\bibitem [{\citenamefont {Shiraishi}\ \emph {et~al.}(2016)\citenamefont {Shiraishi}, \citenamefont {Saito},\ and\ \citenamefont {Tasaki}}]{shiraishi2016universal}%
  \BibitemOpen
  \bibfield  {author} {\bibinfo {author} {\bibfnamefont {N.}~\bibnamefont {Shiraishi}}, \bibinfo {author} {\bibfnamefont {K.}~\bibnamefont {Saito}},\ and\ \bibinfo {author} {\bibfnamefont {H.}~\bibnamefont {Tasaki}},\ }\bibfield  {title} {\bibinfo {title} {Universal trade-off relation between power and efficiency for heat engines},\ }\href {https://doi.org/10.1103/PhysRevLett.117.190601} {\bibfield  {journal} {\bibinfo  {journal} {Phys. Rev. Lett.}\ }\textbf {\bibinfo {volume} {117}},\ \bibinfo {pages} {190601} (\bibinfo {year} {2016})}\BibitemShut {NoStop}%
\bibitem [{\citenamefont {Holubec}\ and\ \citenamefont {Ryabov}(2016)}]{holubecMaximumEfficiencyLowdissipation2016}%
  \BibitemOpen
  \bibfield  {author} {\bibinfo {author} {\bibfnamefont {V.}~\bibnamefont {Holubec}}\ and\ \bibinfo {author} {\bibfnamefont {A.}~\bibnamefont {Ryabov}},\ }\bibfield  {title} {\bibinfo {title} {Maximum efficiency of low-dissipation heat engines at arbitrary power},\ }\href {https://doi.org/10.1088/1742-5468/2016/07/073204} {\bibfield  {journal} {\bibinfo  {journal} {J. Stat. Mech.-Theory Exp.}\ }\textbf {\bibinfo {volume} {2016}},\ \bibinfo {pages} {073204} (\bibinfo {year} {2016})}\BibitemShut {NoStop}%
\bibitem [{\citenamefont {Ma}\ \emph {et~al.}(2018)\citenamefont {Ma}, \citenamefont {Xu}, \citenamefont {Dong},\ and\ \citenamefont {Sun}}]{maUniversalConstraintEfficiency2018}%
  \BibitemOpen
  \bibfield  {author} {\bibinfo {author} {\bibfnamefont {Y.-H.}\ \bibnamefont {Ma}}, \bibinfo {author} {\bibfnamefont {D.}~\bibnamefont {Xu}}, \bibinfo {author} {\bibfnamefont {H.}~\bibnamefont {Dong}},\ and\ \bibinfo {author} {\bibfnamefont {C.-P.}\ \bibnamefont {Sun}},\ }\bibfield  {title} {\bibinfo {title} {Universal constraint for efficiency and power of a low-dissipation heat engine},\ }\href {https://doi.org/10.1103/PhysRevE.98.042112} {\bibfield  {journal} {\bibinfo  {journal} {Phys. Rev. E}\ }\textbf {\bibinfo {volume} {98}},\ \bibinfo {pages} {042112} (\bibinfo {year} {2018})}\BibitemShut {NoStop}%
\bibitem [{\citenamefont {Pietzonka}\ and\ \citenamefont {Seifert}(2018)}]{pietzonkaUniversalTradeOffPower2018}%
  \BibitemOpen
  \bibfield  {author} {\bibinfo {author} {\bibfnamefont {P.}~\bibnamefont {Pietzonka}}\ and\ \bibinfo {author} {\bibfnamefont {U.}~\bibnamefont {Seifert}},\ }\bibfield  {title} {\bibinfo {title} {Universal {{Trade-Off}} between {{Power}}, {{Efficiency}}, and {{Constancy}} in {{Steady-State Heat Engines}}},\ }\href {https://doi.org/10.1103/PhysRevLett.120.190602} {\bibfield  {journal} {\bibinfo  {journal} {Phys. Rev. Lett.}\ }\textbf {\bibinfo {volume} {120}},\ \bibinfo {pages} {190602} (\bibinfo {year} {2018})}\BibitemShut {NoStop}%
\bibitem [{\citenamefont {Yuan}\ \emph {et~al.}(2022)\citenamefont {Yuan}, \citenamefont {Ma},\ and\ \citenamefont {Sun}}]{yuanOptimizingThermodynamicCycles2022}%
  \BibitemOpen
  \bibfield  {author} {\bibinfo {author} {\bibfnamefont {H.}~\bibnamefont {Yuan}}, \bibinfo {author} {\bibfnamefont {Y.-H.}\ \bibnamefont {Ma}},\ and\ \bibinfo {author} {\bibfnamefont {C.~P.}\ \bibnamefont {Sun}},\ }\bibfield  {title} {\bibinfo {title} {Optimizing thermodynamic cycles with two finite-sized reservoirs},\ }\href {https://doi.org/10.1103/PhysRevE.105.L022101} {\bibfield  {journal} {\bibinfo  {journal} {Phys. Rev. E}\ }\textbf {\bibinfo {volume} {105}},\ \bibinfo {pages} {L022101} (\bibinfo {year} {2022})}\BibitemShut {NoStop}%
\bibitem [{\citenamefont {Zhou}\ \emph {et~al.}(2024)\citenamefont {Zhou}, \citenamefont {Ma},\ and\ \citenamefont {Sun}}]{zhou2024finite}%
  \BibitemOpen
  \bibfield  {author} {\bibinfo {author} {\bibfnamefont {T.-J.}\ \bibnamefont {Zhou}}, \bibinfo {author} {\bibfnamefont {Y.-H.}\ \bibnamefont {Ma}},\ and\ \bibinfo {author} {\bibfnamefont {C.~P.}\ \bibnamefont {Sun}},\ }\bibfield  {title} {\bibinfo {title} {Finite-time optimization of a quantum szilard heat engine},\ }\href {https://doi.org/10.1103/PhysRevResearch.6.043001} {\bibfield  {journal} {\bibinfo  {journal} {Phys. Rev. Res.}\ }\textbf {\bibinfo {volume} {6}},\ \bibinfo {pages} {043001} (\bibinfo {year} {2024})}\BibitemShut {NoStop}%
\bibitem [{\citenamefont {Zhao}\ and\ \citenamefont {Ma}(2025)}]{zhao2025revisiting}%
  \BibitemOpen
  \bibfield  {author} {\bibinfo {author} {\bibfnamefont {X.-H.}\ \bibnamefont {Zhao}}\ and\ \bibinfo {author} {\bibfnamefont {Y.-H.}\ \bibnamefont {Ma}},\ }\bibfield  {title} {\bibinfo {title} {Revisiting {{Endoreversible Carnot Engine}}: {{Extending}} the {{Yvon Engine}}},\ }\href {https://doi.org/10.3390/e27020195} {\bibfield  {journal} {\bibinfo  {journal} {Entropy}\ }\textbf {\bibinfo {volume} {27}},\ \bibinfo {pages} {195} (\bibinfo {year} {2025})}\BibitemShut {NoStop}%
\bibitem [{\citenamefont {Brandner}\ and\ \citenamefont {Saito}(2020)}]{brandnerThermodynamicGeometryMicroscopic2020}%
  \BibitemOpen
  \bibfield  {author} {\bibinfo {author} {\bibfnamefont {K.}~\bibnamefont {Brandner}}\ and\ \bibinfo {author} {\bibfnamefont {K.}~\bibnamefont {Saito}},\ }\bibfield  {title} {\bibinfo {title} {Thermodynamic {{Geometry}} of {{Microscopic Heat Engines}}},\ }\href {https://doi.org/10.1103/PhysRevLett.124.040602} {\bibfield  {journal} {\bibinfo  {journal} {Phys. Rev. Lett.}\ }\textbf {\bibinfo {volume} {124}},\ \bibinfo {pages} {040602} (\bibinfo {year} {2020})}\BibitemShut {NoStop}%
\bibitem [{\citenamefont {Ma}\ \emph {et~al.}(2020)\citenamefont {Ma}, \citenamefont {Zhai}, \citenamefont {Chen}, \citenamefont {Sun},\ and\ \citenamefont {Dong}}]{maExperimentalTestTau2020}%
  \BibitemOpen
  \bibfield  {author} {\bibinfo {author} {\bibfnamefont {Y.-H.}\ \bibnamefont {Ma}}, \bibinfo {author} {\bibfnamefont {R.-X.}\ \bibnamefont {Zhai}}, \bibinfo {author} {\bibfnamefont {J.}~\bibnamefont {Chen}}, \bibinfo {author} {\bibfnamefont {C.~P.}\ \bibnamefont {Sun}},\ and\ \bibinfo {author} {\bibfnamefont {H.}~\bibnamefont {Dong}},\ }\bibfield  {title} {\bibinfo {title} {Experimental {{Test}} of the $1/\tau$-{{Scaling Entropy Generation}} in {{Finite-Time Thermodynamics}}},\ }\href {https://doi.org/10.1103/PhysRevLett.125.210601} {\bibfield  {journal} {\bibinfo  {journal} {Phys. Rev. Lett.}\ }\textbf {\bibinfo {volume} {125}},\ \bibinfo {pages} {210601} (\bibinfo {year} {2020})}\BibitemShut {NoStop}%
\bibitem [{\citenamefont {Zhai}\ \emph {et~al.}(2023)\citenamefont {Zhai}, \citenamefont {Cui}, \citenamefont {Ma}, \citenamefont {Sun},\ and\ \citenamefont {Dong}}]{zhaiExperimentalTestPowerefficiency2023}%
  \BibitemOpen
  \bibfield  {author} {\bibinfo {author} {\bibfnamefont {R.-X.}\ \bibnamefont {Zhai}}, \bibinfo {author} {\bibfnamefont {F.-M.}\ \bibnamefont {Cui}}, \bibinfo {author} {\bibfnamefont {Y.-H.}\ \bibnamefont {Ma}}, \bibinfo {author} {\bibfnamefont {C.~P.}\ \bibnamefont {Sun}},\ and\ \bibinfo {author} {\bibfnamefont {H.}~\bibnamefont {Dong}},\ }\bibfield  {title} {\bibinfo {title} {Experimental test of power-efficiency trade-off in a finite-time {{Carnot}} cycle},\ }\href {https://doi.org/10.1103/PhysRevE.107.L042101} {\bibfield  {journal} {\bibinfo  {journal} {Phys. Rev. E}\ }\textbf {\bibinfo {volume} {107}},\ \bibinfo {pages} {L042101} (\bibinfo {year} {2023})}\BibitemShut {NoStop}%
\bibitem [{\citenamefont {Ma}\ \emph {et~al.}(2022)\citenamefont {Ma}, \citenamefont {Chen}, \citenamefont {Sun},\ and\ \citenamefont {Dong}}]{maMinimalEnergyCost2022}%
  \BibitemOpen
  \bibfield  {author} {\bibinfo {author} {\bibfnamefont {Y.-H.}\ \bibnamefont {Ma}}, \bibinfo {author} {\bibfnamefont {J.-F.}\ \bibnamefont {Chen}}, \bibinfo {author} {\bibfnamefont {C.~P.}\ \bibnamefont {Sun}},\ and\ \bibinfo {author} {\bibfnamefont {H.}~\bibnamefont {Dong}},\ }\bibfield  {title} {\bibinfo {title} {Minimal energy cost to initialize a bit with tolerable error},\ }\href {https://doi.org/10.1103/PhysRevE.106.034112} {\bibfield  {journal} {\bibinfo  {journal} {Phys. Rev. E}\ }\textbf {\bibinfo {volume} {106}},\ \bibinfo {pages} {034112} (\bibinfo {year} {2022})}\BibitemShut {NoStop}%
\bibitem [{\citenamefont {Chen}\ \emph {et~al.}(2023)\citenamefont {Chen}, \citenamefont {Zhai}, \citenamefont {Sun},\ and\ \citenamefont {Dong}}]{chen2023geodesic}%
  \BibitemOpen
  \bibfield  {author} {\bibinfo {author} {\bibfnamefont {J.-F.}\ \bibnamefont {Chen}}, \bibinfo {author} {\bibfnamefont {R.-X.}\ \bibnamefont {Zhai}}, \bibinfo {author} {\bibfnamefont {C.~P.}\ \bibnamefont {Sun}},\ and\ \bibinfo {author} {\bibfnamefont {H.}~\bibnamefont {Dong}},\ }\bibfield  {title} {\bibinfo {title} {Geodesic lower bound of the energy consumption to achieve membrane separation within finite time},\ }\href {https://doi.org/10.1103/PRXEnergy.2.033003} {\bibfield  {journal} {\bibinfo  {journal} {PRX Energy}\ }\textbf {\bibinfo {volume} {2}},\ \bibinfo {pages} {033003} (\bibinfo {year} {2023})}\BibitemShut {NoStop}%
\bibitem [{\citenamefont {Zhao}\ \emph {et~al.}(2024)\citenamefont {Zhao}, \citenamefont {Tu},\ and\ \citenamefont {Ma}}]{zhaoEngineeringRatchetbasedParticle2024}%
  \BibitemOpen
  \bibfield  {author} {\bibinfo {author} {\bibfnamefont {X.-H.}\ \bibnamefont {Zhao}}, \bibinfo {author} {\bibfnamefont {Z.~C.}\ \bibnamefont {Tu}},\ and\ \bibinfo {author} {\bibfnamefont {Y.-H.}\ \bibnamefont {Ma}},\ }\bibfield  {title} {\bibinfo {title} {Engineering ratchet-based particle separation via extended shortcuts to isothermality},\ }\href {https://doi.org/10.1103/PhysRevE.110.034105} {\bibfield  {journal} {\bibinfo  {journal} {Phys. Rev. E}\ }\textbf {\bibinfo {volume} {110}},\ \bibinfo {pages} {034105} (\bibinfo {year} {2024})}\BibitemShut {NoStop}%
\bibitem [{\citenamefont {Lei}\ \emph {et~al.}(2025)\citenamefont {Lei}, \citenamefont {Lin}, \citenamefont {Ou},\ and\ \citenamefont {Ma}}]{lei2025universal}%
  \BibitemOpen
  \bibfield  {author} {\bibinfo {author} {\bibfnamefont {J.-R.}\ \bibnamefont {Lei}}, \bibinfo {author} {\bibfnamefont {Y.-Q.}\ \bibnamefont {Lin}}, \bibinfo {author} {\bibfnamefont {S.-G.}\ \bibnamefont {Ou}},\ and\ \bibinfo {author} {\bibfnamefont {Y.-H.}\ \bibnamefont {Ma}},\ }\bibfield  {title} {\bibinfo {title} {Universal power-efficiency trade-off in battery charging},\ }\href {https://doi.org/10.15302/frontphys.2025.042202} {\bibfield  {journal} {\bibinfo  {journal} {Front. Phys.}\ }\textbf {\bibinfo {volume} {20}},\ \bibinfo {pages} {042202} (\bibinfo {year} {2025})}\BibitemShut {NoStop}%
\bibitem [{\citenamefont {Zhai}\ and\ \citenamefont {Dong}(2025)}]{zhai2025power}%
  \BibitemOpen
  \bibfield  {author} {\bibinfo {author} {\bibfnamefont {R.-X.}\ \bibnamefont {Zhai}}\ and\ \bibinfo {author} {\bibfnamefont {H.}~\bibnamefont {Dong}},\ }\bibfield  {title} {\bibinfo {title} {Power-efficiency constraint for chemical motors},\ }\href {https://doi.org/10.1103/PhysRevE.111.024404} {\bibfield  {journal} {\bibinfo  {journal} {Phys. Rev. E}\ }\textbf {\bibinfo {volume} {111}},\ \bibinfo {pages} {024404} (\bibinfo {year} {2025})}\BibitemShut {NoStop}%
\bibitem [{\citenamefont {Curzon}\ and\ \citenamefont {Ahlborn}(1975)}]{curzonEfficiencyCarnotEngine1975}%
  \BibitemOpen
  \bibfield  {author} {\bibinfo {author} {\bibfnamefont {F.~L.}\ \bibnamefont {Curzon}}\ and\ \bibinfo {author} {\bibfnamefont {B.}~\bibnamefont {Ahlborn}},\ }\bibfield  {title} {\bibinfo {title} {Efficiency of a carnot engine at maximum power output},\ }\href {https://doi.org/10.1119/1.10023} {\bibfield  {journal} {\bibinfo  {journal} {Am. J. Phys.}\ }\textbf {\bibinfo {volume} {43}},\ \bibinfo {pages} {22} (\bibinfo {year} {1975})}\BibitemShut {NoStop}%
\bibitem [{\citenamefont {Tu}(2008)}]{tuEfficiencyMaximumPower2008}%
  \BibitemOpen
  \bibfield  {author} {\bibinfo {author} {\bibfnamefont {Z.~C.}\ \bibnamefont {Tu}},\ }\bibfield  {title} {\bibinfo {title} {Efficiency at maximum power of feynman's ratchet as a heat engine},\ }\href {https://doi.org/10.1088/1751-8113/41/31/312003} {\bibfield  {journal} {\bibinfo  {journal} {J. Phys. A: Math. Theor.}\ }\textbf {\bibinfo {volume} {41}},\ \bibinfo {pages} {312003} (\bibinfo {year} {2008})}\BibitemShut {NoStop}%
\bibitem [{\citenamefont {{Van den Broeck}}(2005)}]{vandenbroeckThermodynamicEfficiencyMaximum2005}%
  \BibitemOpen
  \bibfield  {author} {\bibinfo {author} {\bibfnamefont {C.}~\bibnamefont {{Van den Broeck}}},\ }\bibfield  {title} {\bibinfo {title} {Thermodynamic efficiency at maximum power},\ }\href {https://doi.org/10.1103/PhysRevLett.95.190602} {\bibfield  {journal} {\bibinfo  {journal} {Phys. Rev. Lett.}\ }\textbf {\bibinfo {volume} {95}},\ \bibinfo {pages} {190602} (\bibinfo {year} {2005})}\BibitemShut {NoStop}%
\bibitem [{\citenamefont {Esposito}\ \emph {et~al.}(2010)\citenamefont {Esposito}, \citenamefont {Kawai}, \citenamefont {Lindenberg},\ and\ \citenamefont {{Van den Broeck}}}]{espositoEfficiencyMaximumPower2010}%
  \BibitemOpen
  \bibfield  {author} {\bibinfo {author} {\bibfnamefont {M.}~\bibnamefont {Esposito}}, \bibinfo {author} {\bibfnamefont {R.}~\bibnamefont {Kawai}}, \bibinfo {author} {\bibfnamefont {K.}~\bibnamefont {Lindenberg}},\ and\ \bibinfo {author} {\bibfnamefont {C.}~\bibnamefont {{Van den Broeck}}},\ }\bibfield  {title} {\bibinfo {title} {Efficiency at maximum power of low-dissipation carnot engines},\ }\href {https://doi.org/10.1103/PhysRevLett.105.150603} {\bibfield  {journal} {\bibinfo  {journal} {Phys. Rev. Lett.}\ }\textbf {\bibinfo {volume} {105}},\ \bibinfo {pages} {150603} (\bibinfo {year} {2010})}\BibitemShut {NoStop}%
\bibitem [{\citenamefont {Blanchard}(1980)}]{blanchardCoefficientPerformanceFinite1980}%
  \BibitemOpen
  \bibfield  {author} {\bibinfo {author} {\bibfnamefont {C.~H.}\ \bibnamefont {Blanchard}},\ }\bibfield  {title} {\bibinfo {title} {Coefficient of performance for finite speed heat pump},\ }\href {https://doi.org/10.1063/1.328020} {\bibfield  {journal} {\bibinfo  {journal} {J. Appl. Phys.}\ }\textbf {\bibinfo {volume} {51}},\ \bibinfo {pages} {2471} (\bibinfo {year} {1980})}\BibitemShut {NoStop}%
\bibitem [{\citenamefont {Chen}\ \emph {et~al.}(1996)\citenamefont {Chen}, \citenamefont {Sun},\ and\ \citenamefont {Wu}}]{Chen1996TheIO}%
  \BibitemOpen
  \bibfield  {author} {\bibinfo {author} {\bibfnamefont {L.~G.}\ \bibnamefont {Chen}}, \bibinfo {author} {\bibfnamefont {F.}~\bibnamefont {Sun}},\ and\ \bibinfo {author} {\bibfnamefont {C.~S.}\ \bibnamefont {Wu}},\ }\bibfield  {title} {\bibinfo {title} {The influence of heat-transfer law on the endo-reversible carnot refrigerator},\ }\href {https://api.semanticscholar.org/CorpusID:102458148} {\bibfield  {journal} {\bibinfo  {journal} {J. I. Energy}\ }\textbf {\bibinfo {volume} {69}},\ \bibinfo {pages} {96} (\bibinfo {year} {1996})}\BibitemShut {NoStop}%
\bibitem [{\citenamefont {Wang}\ \emph {et~al.}(2012)\citenamefont {Wang}, \citenamefont {Li}, \citenamefont {Tu}, \citenamefont {Hern{\'a}ndez},\ and\ \citenamefont {Roco}}]{wangCoefficientPerformanceMaximum2012}%
  \BibitemOpen
  \bibfield  {author} {\bibinfo {author} {\bibfnamefont {Y.}~\bibnamefont {Wang}}, \bibinfo {author} {\bibfnamefont {M.}~\bibnamefont {Li}}, \bibinfo {author} {\bibfnamefont {Z.~C.}\ \bibnamefont {Tu}}, \bibinfo {author} {\bibfnamefont {A.~C.}\ \bibnamefont {Hern{\'a}ndez}},\ and\ \bibinfo {author} {\bibfnamefont {J.~M.~M.}\ \bibnamefont {Roco}},\ }\bibfield  {title} {\bibinfo {title} {Coefficient of performance at maximum figure of merit and its bounds for low-dissipation {{Carnot-like}} refrigerators},\ }\href {https://doi.org/10.1103/PhysRevE.86.011127} {\bibfield  {journal} {\bibinfo  {journal} {Phys. Rev. E}\ }\textbf {\bibinfo {volume} {86}},\ \bibinfo {pages} {011127} (\bibinfo {year} {2012})}\BibitemShut {NoStop}%
\bibitem [{\citenamefont {Izumida}\ \emph {et~al.}(2013)\citenamefont {Izumida}, \citenamefont {Okuda}, \citenamefont {Calvo~Hern{\'a}ndez},\ and\ \citenamefont {Roco}}]{izumidaCoefficientPerformanceOptimized2013}%
  \BibitemOpen
  \bibfield  {author} {\bibinfo {author} {\bibfnamefont {Y.}~\bibnamefont {Izumida}}, \bibinfo {author} {\bibfnamefont {K.}~\bibnamefont {Okuda}}, \bibinfo {author} {\bibfnamefont {A.}~\bibnamefont {Calvo~Hern{\'a}ndez}},\ and\ \bibinfo {author} {\bibfnamefont {J.~M.~M.}\ \bibnamefont {Roco}},\ }\bibfield  {title} {\bibinfo {title} {Coefficient of performance under optimized figure of merit in minimally nonlinear irreversible refrigerator},\ }\href {https://doi.org/10.1209/0295-5075/101/10005} {\bibfield  {journal} {\bibinfo  {journal} {Europhys. Lett.}\ }\textbf {\bibinfo {volume} {101}},\ \bibinfo {pages} {10005} (\bibinfo {year} {2013})}\BibitemShut {NoStop}%
\bibitem [{\citenamefont {Holubec}\ and\ \citenamefont {Ye}(2020)}]{holubecMaximumEfficiencyLowdissipation2020}%
  \BibitemOpen
  \bibfield  {author} {\bibinfo {author} {\bibfnamefont {V.}~\bibnamefont {Holubec}}\ and\ \bibinfo {author} {\bibfnamefont {Z.}~\bibnamefont {Ye}},\ }\bibfield  {title} {\bibinfo {title} {Maximum efficiency of low-dissipation refrigerators at arbitrary cooling power},\ }\href {https://doi.org/10.1103/PhysRevE.101.052124} {\bibfield  {journal} {\bibinfo  {journal} {Phys. Rev. E}\ }\textbf {\bibinfo {volume} {101}},\ \bibinfo {pages} {052124} (\bibinfo {year} {2020})}\BibitemShut {NoStop}%
\bibitem [{\citenamefont {Ye}\ and\ \citenamefont {Holubec}(2022)}]{yeMaximumEfficiencyLowdissipation2022}%
  \BibitemOpen
  \bibfield  {author} {\bibinfo {author} {\bibfnamefont {Z.}~\bibnamefont {Ye}}\ and\ \bibinfo {author} {\bibfnamefont {V.}~\bibnamefont {Holubec}},\ }\bibfield  {title} {\bibinfo {title} {Maximum efficiency of low-dissipation heat pumps at given heating load},\ }\href {https://doi.org/10.1103/PhysRevE.105.024139} {\bibfield  {journal} {\bibinfo  {journal} {Phys. Rev. E}\ }\textbf {\bibinfo {volume} {105}},\ \bibinfo {pages} {024139} (\bibinfo {year} {2022})}\BibitemShut {NoStop}%
\bibitem [{\citenamefont {Chen}\ \emph {et~al.}(1995)\citenamefont {Chen}, \citenamefont {Sun}, \citenamefont {Cheng},\ and\ \citenamefont {Chen}}]{chenStudyOptimalPerformance1995}%
  \BibitemOpen
  \bibfield  {author} {\bibinfo {author} {\bibfnamefont {W.~Z.}\ \bibnamefont {Chen}}, \bibinfo {author} {\bibfnamefont {F.~R.}\ \bibnamefont {Sun}}, \bibinfo {author} {\bibfnamefont {S.~M.}\ \bibnamefont {Cheng}},\ and\ \bibinfo {author} {\bibfnamefont {L.~G.}\ \bibnamefont {Chen}},\ }\bibfield  {title} {\bibinfo {title} {Study on optimal performance and working temperatures of endoreversible forward and reverse carnot cycles},\ }\href {https://doi.org/10.1002/er.4440190903} {\bibfield  {journal} {\bibinfo  {journal} {Int. J. Energ. Res.}\ }\textbf {\bibinfo {volume} {19}},\ \bibinfo {pages} {751} (\bibinfo {year} {1995})}\BibitemShut {NoStop}%
\bibitem [{\citenamefont {De~Tom{\'a}s}\ \emph {et~al.}(2013)\citenamefont {De~Tom{\'a}s}, \citenamefont {Roco}, \citenamefont {Hern{\'a}ndez}, \citenamefont {Wang},\ and\ \citenamefont {Tu}}]{de2013low}%
  \BibitemOpen
  \bibfield  {author} {\bibinfo {author} {\bibfnamefont {C.}~\bibnamefont {De~Tom{\'a}s}}, \bibinfo {author} {\bibfnamefont {J.}~\bibnamefont {Roco}}, \bibinfo {author} {\bibfnamefont {A.~C.}\ \bibnamefont {Hern{\'a}ndez}}, \bibinfo {author} {\bibfnamefont {Y.}~\bibnamefont {Wang}},\ and\ \bibinfo {author} {\bibfnamefont {Z.}~\bibnamefont {Tu}},\ }\bibfield  {title} {\bibinfo {title} {Low-dissipation heat devices: Unified trade-off optimization and bounds},\ }\href {https://doi.org/10.1103/PhysRevE.87.012105} {\bibfield  {journal} {\bibinfo  {journal} {Phys. Rev. E}\ }\textbf {\bibinfo {volume} {87}},\ \bibinfo {pages} {012105} (\bibinfo {year} {2013})}\BibitemShut {NoStop}%
\bibitem [{\citenamefont {Johal}(2019)}]{johalPerformanceOptimizationLowdissipation2019}%
  \BibitemOpen
  \bibfield  {author} {\bibinfo {author} {\bibfnamefont {R.~S.}\ \bibnamefont {Johal}},\ }\bibfield  {title} {\bibinfo {title} {Performance optimization of low-dissipation thermal machines revisited},\ }\href {https://doi.org/10.1103/PhysRevE.100.052101} {\bibfield  {journal} {\bibinfo  {journal} {Phys. Rev. E}\ }\textbf {\bibinfo {volume} {100}},\ \bibinfo {pages} {052101} (\bibinfo {year} {2019})}\BibitemShut {NoStop}%
\bibitem [{\citenamefont {{Gonzalez-Ayala}}\ \emph {et~al.}(2018)\citenamefont {{Gonzalez-Ayala}}, \citenamefont {Medina}, \citenamefont {Roco},\ and\ \citenamefont {Hern{\'a}ndez}}]{gonzalez-ayalaEntropyGenerationUnified2018}%
  \BibitemOpen
  \bibfield  {author} {\bibinfo {author} {\bibfnamefont {J.}~\bibnamefont {{Gonzalez-Ayala}}}, \bibinfo {author} {\bibfnamefont {A.}~\bibnamefont {Medina}}, \bibinfo {author} {\bibfnamefont {J.~M.~M.}\ \bibnamefont {Roco}},\ and\ \bibinfo {author} {\bibfnamefont {A.~C.}\ \bibnamefont {Hern{\'a}ndez}},\ }\bibfield  {title} {\bibinfo {title} {Entropy generation and unified optimization of {{Carnot-like}} and low-dissipation refrigerators},\ }\href {https://doi.org/10.1103/PhysRevE.97.022139} {\bibfield  {journal} {\bibinfo  {journal} {Phys. Rev. E}\ }\textbf {\bibinfo {volume} {97}},\ \bibinfo {pages} {022139} (\bibinfo {year} {2018})}\BibitemShut {NoStop}%
\bibitem [{\citenamefont {Leff}\ and\ \citenamefont {Teeters}(1978)}]{leffEERCOPSecond1978}%
  \BibitemOpen
  \bibfield  {author} {\bibinfo {author} {\bibfnamefont {H.~S.}\ \bibnamefont {Leff}}\ and\ \bibinfo {author} {\bibfnamefont {W.~D.}\ \bibnamefont {Teeters}},\ }\bibfield  {title} {\bibinfo {title} {{{EER}}, {{COP}}, and the second law efficiency for air conditioners},\ }\href {https://doi.org/10.1119/1.11174} {\bibfield  {journal} {\bibinfo  {journal} {Am. J. Phys.}\ }\textbf {\bibinfo {volume} {46}},\ \bibinfo {pages} {19} (\bibinfo {year} {1978})}\BibitemShut {NoStop}%
\bibitem [{\citenamefont {Gordon}\ \emph {et~al.}(1997)\citenamefont {Gordon}, \citenamefont {Ng},\ and\ \citenamefont {Chua}}]{gordonOptimizingChillerOperation1997}%
  \BibitemOpen
  \bibfield  {author} {\bibinfo {author} {\bibfnamefont {J.}~\bibnamefont {Gordon}}, \bibinfo {author} {\bibfnamefont {K.}~\bibnamefont {Ng}},\ and\ \bibinfo {author} {\bibfnamefont {H.}~\bibnamefont {Chua}},\ }\bibfield  {title} {\bibinfo {title} {Optimizing chiller operation based on finite-time thermodynamics: Universal modeling and experimental confirmation},\ }\href {https://doi.org/10.1016/S0140-7007(96)00074-6} {\bibfield  {journal} {\bibinfo  {journal} {Int. J. Refrig.}\ }\textbf {\bibinfo {volume} {20}},\ \bibinfo {pages} {191} (\bibinfo {year} {1997})}\BibitemShut {NoStop}%
\bibitem [{\citenamefont {Wang}\ and\ \citenamefont {Tu}(2012)}]{wang2012efficiency}%
  \BibitemOpen
  \bibfield  {author} {\bibinfo {author} {\bibfnamefont {Y.}~\bibnamefont {Wang}}\ and\ \bibinfo {author} {\bibfnamefont {Z.}~\bibnamefont {Tu}},\ }\bibfield  {title} {\bibinfo {title} {Efficiency at maximum power output of linear irreversible carnot-like heat engines},\ }\href {https://doi.org/https://doi.org/10.1103/PhysRevE.85.011127} {\bibfield  {journal} {\bibinfo  {journal} {Phys. Rev. E}\ }\textbf {\bibinfo {volume} {85}},\ \bibinfo {pages} {011127} (\bibinfo {year} {2012})}\BibitemShut {NoStop}%
\bibitem [{\citenamefont {Johal}(2017)}]{johal2017heat}%
  \BibitemOpen
  \bibfield  {author} {\bibinfo {author} {\bibfnamefont {R.~S.}\ \bibnamefont {Johal}},\ }\bibfield  {title} {\bibinfo {title} {Heat engines at optimal power: Low-dissipation versus endoreversible model},\ }\href {https://doi.org/https://doi.org/10.1103/PhysRevE.96.012151} {\bibfield  {journal} {\bibinfo  {journal} {Phys. Rev. E}\ }\textbf {\bibinfo {volume} {96}},\ \bibinfo {pages} {012151} (\bibinfo {year} {2017})}\BibitemShut {NoStop}%
\bibitem [{\citenamefont {Esposito}\ \emph {et~al.}(2009)\citenamefont {Esposito}, \citenamefont {Lindenberg},\ and\ \citenamefont {{Van den Broeck}}}]{espositoUniversalityEfficiencyMaximum2009}%
  \BibitemOpen
  \bibfield  {author} {\bibinfo {author} {\bibfnamefont {M.}~\bibnamefont {Esposito}}, \bibinfo {author} {\bibfnamefont {K.}~\bibnamefont {Lindenberg}},\ and\ \bibinfo {author} {\bibfnamefont {C.}~\bibnamefont {{Van den Broeck}}},\ }\bibfield  {title} {\bibinfo {title} {Universality of {{Efficiency}} at {{Maximum Power}}},\ }\href {https://doi.org/10.1103/PhysRevLett.102.130602} {\bibfield  {journal} {\bibinfo  {journal} {Phys. Rev. Lett.}\ }\textbf {\bibinfo {volume} {102}},\ \bibinfo {pages} {130602} (\bibinfo {year} {2009})}\BibitemShut {NoStop}%
\bibitem [{\citenamefont {Chen}\ \emph {et~al.}(2019{\natexlab{a}})\citenamefont {Chen}, \citenamefont {Sun},\ and\ \citenamefont {Dong}}]{chen2019achieve}%
  \BibitemOpen
  \bibfield  {author} {\bibinfo {author} {\bibfnamefont {J.-F.}\ \bibnamefont {Chen}}, \bibinfo {author} {\bibfnamefont {C.-P.}\ \bibnamefont {Sun}},\ and\ \bibinfo {author} {\bibfnamefont {H.}~\bibnamefont {Dong}},\ }\bibfield  {title} {\bibinfo {title} {Achieve higher efficiency at maximum power with finite-time quantum otto cycle},\ }\href {https://doi.org/10.1103/PhysRevE.100.062140} {\bibfield  {journal} {\bibinfo  {journal} {Phys. Rev. E}\ }\textbf {\bibinfo {volume} {100}},\ \bibinfo {pages} {062140} (\bibinfo {year} {2019}{\natexlab{a}})}\BibitemShut {NoStop}%
\bibitem [{\citenamefont {Fei}\ \emph {et~al.}(2022)\citenamefont {Fei}, \citenamefont {Chen},\ and\ \citenamefont {Ma}}]{fei2022efficiency}%
  \BibitemOpen
  \bibfield  {author} {\bibinfo {author} {\bibfnamefont {Z.}~\bibnamefont {Fei}}, \bibinfo {author} {\bibfnamefont {J.-F.}\ \bibnamefont {Chen}},\ and\ \bibinfo {author} {\bibfnamefont {Y.-H.}\ \bibnamefont {Ma}},\ }\bibfield  {title} {\bibinfo {title} {Efficiency statistics of a quantum otto cycle},\ }\href {https://doi.org/10.1103/PhysRevA.105.022609} {\bibfield  {journal} {\bibinfo  {journal} {Phys. Rev. A}\ }\textbf {\bibinfo {volume} {105}},\ \bibinfo {pages} {022609} (\bibinfo {year} {2022})}\BibitemShut {NoStop}%
\bibitem [{\citenamefont {Liang}\ \emph {et~al.}(2025)\citenamefont {Liang}, \citenamefont {Ma}, \citenamefont {Busiello},\ and\ \citenamefont {De~Los~Rios}}]{liang2023}%
  \BibitemOpen
  \bibfield  {author} {\bibinfo {author} {\bibfnamefont {S.}~\bibnamefont {Liang}}, \bibinfo {author} {\bibfnamefont {Y.-H.}\ \bibnamefont {Ma}}, \bibinfo {author} {\bibfnamefont {D.~M.}\ \bibnamefont {Busiello}},\ and\ \bibinfo {author} {\bibfnamefont {P.}~\bibnamefont {De~Los~Rios}},\ }\bibfield  {title} {\bibinfo {title} {Minimal {{Model}} for {{Carnot Efficiency}} at {{Maximum Power}}},\ }\href {https://doi.org/10.1103/PhysRevLett.134.027101} {\bibfield  {journal} {\bibinfo  {journal} {Phys. Rev. Lett.}\ }\textbf {\bibinfo {volume} {134}},\ \bibinfo {pages} {027101} (\bibinfo {year} {2025})}\BibitemShut {NoStop}%
\bibitem [{\citenamefont {Pancotti}\ \emph {et~al.}(2020)\citenamefont {Pancotti}, \citenamefont {Scandi}, \citenamefont {Mitchison},\ and\ \citenamefont {Perarnau-Llobet}}]{pancotti2020speed}%
  \BibitemOpen
  \bibfield  {author} {\bibinfo {author} {\bibfnamefont {N.}~\bibnamefont {Pancotti}}, \bibinfo {author} {\bibfnamefont {M.}~\bibnamefont {Scandi}}, \bibinfo {author} {\bibfnamefont {M.~T.}\ \bibnamefont {Mitchison}},\ and\ \bibinfo {author} {\bibfnamefont {M.}~\bibnamefont {Perarnau-Llobet}},\ }\bibfield  {title} {\bibinfo {title} {Speed-ups to isothermality: Enhanced quantum thermal machines through control of the system-bath coupling},\ }\href {https://doi.org/10.1103/PhysRevX.10.031015} {\bibfield  {journal} {\bibinfo  {journal} {Phys. Rev. X}\ }\textbf {\bibinfo {volume} {10}},\ \bibinfo {pages} {031015} (\bibinfo {year} {2020})}\BibitemShut {NoStop}%
\bibitem [{\citenamefont {Wang}\ and\ \citenamefont {Tu}(2013)}]{yang2013bounds}%
  \BibitemOpen
  \bibfield  {author} {\bibinfo {author} {\bibfnamefont {Y.}~\bibnamefont {Wang}}\ and\ \bibinfo {author} {\bibfnamefont {Z.~C.}\ \bibnamefont {Tu}},\ }\bibfield  {title} {\bibinfo {title} {Bounds of efficiency at maximum power for normal-, sub-and super-dissipative carnot-like heat engines},\ }\href {https://doi.org/10.1088/0253-6102/59/2/08} {\bibfield  {journal} {\bibinfo  {journal} {Commun. Theor. Phys.}\ }\textbf {\bibinfo {volume} {59}},\ \bibinfo {pages} {175} (\bibinfo {year} {2013})}\BibitemShut {NoStop}%
\bibitem [{\citenamefont {Amelkin}\ \emph {et~al.}(2004)\citenamefont {Amelkin}, \citenamefont {Andresen}, \citenamefont {Burzler}, \citenamefont {Hoffmann},\ and\ \citenamefont {Tsirlin}}]{amelkinMaximumPowerProcesses2004}%
  \BibitemOpen
  \bibfield  {author} {\bibinfo {author} {\bibfnamefont {S.~A.}\ \bibnamefont {Amelkin}}, \bibinfo {author} {\bibfnamefont {B.}~\bibnamefont {Andresen}}, \bibinfo {author} {\bibfnamefont {J.~M.}\ \bibnamefont {Burzler}}, \bibinfo {author} {\bibfnamefont {K.~H.}\ \bibnamefont {Hoffmann}},\ and\ \bibinfo {author} {\bibfnamefont {A.~M.}\ \bibnamefont {Tsirlin}},\ }\bibfield  {title} {\bibinfo {title} {Maximum power processes for multi-source endoreversible heat engines},\ }\href {https://doi.org/10.1088/0022-3727/37/9/015} {\bibfield  {journal} {\bibinfo  {journal} {J. Phys. D: Appl. Phys.}\ }\textbf {\bibinfo {volume} {37}},\ \bibinfo {pages} {1400} (\bibinfo {year} {2004})}\BibitemShut {NoStop}%
\bibitem [{\citenamefont {Salamon}\ and\ \citenamefont {Berry}(1983)}]{salamonThermodynamicLengthDissipated1983}%
  \BibitemOpen
  \bibfield  {author} {\bibinfo {author} {\bibfnamefont {P.}~\bibnamefont {Salamon}}\ and\ \bibinfo {author} {\bibfnamefont {R.~S.}\ \bibnamefont {Berry}},\ }\bibfield  {title} {\bibinfo {title} {Thermodynamic {{Length}} and {{Dissipated Availability}}},\ }\href {https://doi.org/10.1103/PhysRevLett.51.1127} {\bibfield  {journal} {\bibinfo  {journal} {Phys. Rev. Lett.}\ }\textbf {\bibinfo {volume} {51}},\ \bibinfo {pages} {1127} (\bibinfo {year} {1983})}\BibitemShut {NoStop}%
\bibitem [{\citenamefont {Andresen}\ \emph {et~al.}(1988)\citenamefont {Andresen}, \citenamefont {Berry}, \citenamefont {Gilmore}, \citenamefont {Ihrig},\ and\ \citenamefont {Salamon}}]{andresenThermodynamicGeometryMetrics1988}%
  \BibitemOpen
  \bibfield  {author} {\bibinfo {author} {\bibfnamefont {B.}~\bibnamefont {Andresen}}, \bibinfo {author} {\bibfnamefont {R.~S.}\ \bibnamefont {Berry}}, \bibinfo {author} {\bibfnamefont {R.}~\bibnamefont {Gilmore}}, \bibinfo {author} {\bibfnamefont {E.}~\bibnamefont {Ihrig}},\ and\ \bibinfo {author} {\bibfnamefont {P.}~\bibnamefont {Salamon}},\ }\bibfield  {title} {\bibinfo {title} {Thermodynamic geometry and the metrics of {{Weinhold}} and {{Gilmore}}},\ }\href {https://doi.org/10.1103/PhysRevA.37.845} {\bibfield  {journal} {\bibinfo  {journal} {Phys. Rev. A}\ }\textbf {\bibinfo {volume} {37}},\ \bibinfo {pages} {845} (\bibinfo {year} {1988})}\BibitemShut {NoStop}%
\bibitem [{\citenamefont {Hoffmann}\ \emph {et~al.}(1989)\citenamefont {Hoffmann}, \citenamefont {Andresen},\ and\ \citenamefont {Salamon}}]{hoffmannMeasuresDissipation1989}%
  \BibitemOpen
  \bibfield  {author} {\bibinfo {author} {\bibfnamefont {K.~H.}\ \bibnamefont {Hoffmann}}, \bibinfo {author} {\bibfnamefont {B.}~\bibnamefont {Andresen}},\ and\ \bibinfo {author} {\bibfnamefont {P.}~\bibnamefont {Salamon}},\ }\bibfield  {title} {\bibinfo {title} {Measures of dissipation},\ }\href {https://doi.org/10.1103/PhysRevA.39.3618} {\bibfield  {journal} {\bibinfo  {journal} {Phys. Rev. A}\ }\textbf {\bibinfo {volume} {39}},\ \bibinfo {pages} {3618} (\bibinfo {year} {1989})}\BibitemShut {NoStop}%
\bibitem [{\citenamefont {Crooks}(2007)}]{crooksMeasuringThermodynamicLength2007a}%
  \BibitemOpen
  \bibfield  {author} {\bibinfo {author} {\bibfnamefont {G.~E.}\ \bibnamefont {Crooks}},\ }\bibfield  {title} {\bibinfo {title} {Measuring {{Thermodynamic Length}}},\ }\href {https://doi.org/10.1103/PhysRevLett.99.100602} {\bibfield  {journal} {\bibinfo  {journal} {Phys. Rev. Lett.}\ }\textbf {\bibinfo {volume} {99}},\ \bibinfo {pages} {100602} (\bibinfo {year} {2007})}\BibitemShut {NoStop}%
\bibitem [{\citenamefont {Chen}\ \emph {et~al.}(2019{\natexlab{b}})\citenamefont {Chen}, \citenamefont {Sun},\ and\ \citenamefont {Dong}}]{chenBoostingPerformanceQuantum2019}%
  \BibitemOpen
  \bibfield  {author} {\bibinfo {author} {\bibfnamefont {J.-F.}\ \bibnamefont {Chen}}, \bibinfo {author} {\bibfnamefont {C.-P.}\ \bibnamefont {Sun}},\ and\ \bibinfo {author} {\bibfnamefont {H.}~\bibnamefont {Dong}},\ }\bibfield  {title} {\bibinfo {title} {Boosting the performance of quantum {{Otto}} heat engines},\ }\href {https://doi.org/10.1103/PhysRevE.100.032144} {\bibfield  {journal} {\bibinfo  {journal} {Phys. Rev. E}\ }\textbf {\bibinfo {volume} {100}},\ \bibinfo {pages} {032144} (\bibinfo {year} {2019}{\natexlab{b}})}\BibitemShut {NoStop}%
\bibitem [{\citenamefont {Li}\ \emph {et~al.}(2022)\citenamefont {Li}, \citenamefont {Chen}, \citenamefont {Sun},\ and\ \citenamefont {Dong}}]{liGeodesicPathMinimal2022}%
  \BibitemOpen
  \bibfield  {author} {\bibinfo {author} {\bibfnamefont {G.}~\bibnamefont {Li}}, \bibinfo {author} {\bibfnamefont {J.-F.}\ \bibnamefont {Chen}}, \bibinfo {author} {\bibfnamefont {C.~P.}\ \bibnamefont {Sun}},\ and\ \bibinfo {author} {\bibfnamefont {H.}~\bibnamefont {Dong}},\ }\bibfield  {title} {\bibinfo {title} {Geodesic {{Path}} for the {{Minimal Energy Cost}} in {{Shortcuts}} to {{Isothermality}}},\ }\href {https://doi.org/10.1103/PhysRevLett.128.230603} {\bibfield  {journal} {\bibinfo  {journal} {Phys. Rev. Lett.}\ }\textbf {\bibinfo {volume} {128}},\ \bibinfo {pages} {230603} (\bibinfo {year} {2022})}\BibitemShut {NoStop}%
\bibitem [{\citenamefont {Frim}\ and\ \citenamefont {DeWeese}(2022)}]{frimGeometricBoundEfficiency2022}%
  \BibitemOpen
  \bibfield  {author} {\bibinfo {author} {\bibfnamefont {A.~G.}\ \bibnamefont {Frim}}\ and\ \bibinfo {author} {\bibfnamefont {M.~R.}\ \bibnamefont {DeWeese}},\ }\bibfield  {title} {\bibinfo {title} {Geometric {{Bound}} on the {{Efficiency}} of {{Irreversible Thermodynamic Cycles}}},\ }\href {https://doi.org/10.1103/PhysRevLett.128.230601} {\bibfield  {journal} {\bibinfo  {journal} {Phys. Rev. Lett.}\ }\textbf {\bibinfo {volume} {128}},\ \bibinfo {pages} {230601} (\bibinfo {year} {2022})}\BibitemShut {NoStop}%
\bibitem [{\citenamefont {Chen}(2022)}]{chenOptimizingBrownianHeat2022}%
  \BibitemOpen
  \bibfield  {author} {\bibinfo {author} {\bibfnamefont {J.-F.}\ \bibnamefont {Chen}},\ }\bibfield  {title} {\bibinfo {title} {Optimizing brownian heat engine with shortcut strategy},\ }\href {https://doi.org/10.1103/PhysRevE.106.054108} {\bibfield  {journal} {\bibinfo  {journal} {Phys. Rev. E}\ }\textbf {\bibinfo {volume} {106}},\ \bibinfo {pages} {054108} (\bibinfo {year} {2022})}\BibitemShut {NoStop}%
\bibitem [{\citenamefont {Izumida}\ and\ \citenamefont {Okuda}(2014)}]{izumidaWorkOutputEfficiency2014}%
  \BibitemOpen
  \bibfield  {author} {\bibinfo {author} {\bibfnamefont {Y.}~\bibnamefont {Izumida}}\ and\ \bibinfo {author} {\bibfnamefont {K.}~\bibnamefont {Okuda}},\ }\bibfield  {title} {\bibinfo {title} {Work {{Output}} and {{Efficiency}} at {{Maximum Power}} of {{Linear Irreversible Heat Engines Operating}} with a {{Finite-Sized Heat Source}}},\ }\href {https://doi.org/10.1103/PhysRevLett.112.180603} {\bibfield  {journal} {\bibinfo  {journal} {Phys. Rev. Lett.}\ }\textbf {\bibinfo {volume} {112}},\ \bibinfo {pages} {180603} (\bibinfo {year} {2014})}\BibitemShut {NoStop}%
\bibitem [{\citenamefont {Li}\ \emph {et~al.}(2017)\citenamefont {Li}, \citenamefont {Quan},\ and\ \citenamefont {Tu}}]{liShortcutsIsothermalityNonequilibrium2017}%
  \BibitemOpen
  \bibfield  {author} {\bibinfo {author} {\bibfnamefont {G.}~\bibnamefont {Li}}, \bibinfo {author} {\bibfnamefont {H.~T.}\ \bibnamefont {Quan}},\ and\ \bibinfo {author} {\bibfnamefont {Z.~C.}\ \bibnamefont {Tu}},\ }\bibfield  {title} {\bibinfo {title} {Shortcuts to isothermality and nonequilibrium work relations},\ }\href {https://doi.org/10.1103/PhysRevE.96.012144} {\bibfield  {journal} {\bibinfo  {journal} {Phys. Rev. E}\ }\textbf {\bibinfo {volume} {96}},\ \bibinfo {pages} {012144} (\bibinfo {year} {2017})}\BibitemShut {NoStop}%
\bibitem [{\citenamefont {Rolandi}\ \emph {et~al.}(2023)\citenamefont {Rolandi}, \citenamefont {Abiuso},\ and\ \citenamefont {{Perarnau-Llobet}}}]{rolandiCollectiveAdvantagesFiniteTime2023}%
  \BibitemOpen
  \bibfield  {author} {\bibinfo {author} {\bibfnamefont {A.}~\bibnamefont {Rolandi}}, \bibinfo {author} {\bibfnamefont {P.}~\bibnamefont {Abiuso}},\ and\ \bibinfo {author} {\bibfnamefont {M.}~\bibnamefont {{Perarnau-Llobet}}},\ }\bibfield  {title} {\bibinfo {title} {Collective {{Advantages}} in {{Finite-Time Thermodynamics}}},\ }\href {https://doi.org/10.1103/PhysRevLett.131.210401} {\bibfield  {journal} {\bibinfo  {journal} {Phys. Rev. Lett.}\ }\textbf {\bibinfo {volume} {131}},\ \bibinfo {pages} {210401} (\bibinfo {year} {2023})}\BibitemShut {NoStop}%
\bibitem [{\citenamefont {Izumida}(2022)}]{izumida2022irreversible}%
  \BibitemOpen
  \bibfield  {author} {\bibinfo {author} {\bibfnamefont {Y.}~\bibnamefont {Izumida}},\ }\bibfield  {title} {\bibinfo {title} {Irreversible efficiency and carnot theorem for heat engines operating with multiple heat baths in linear response regime},\ }\href {https://doi.org/10.1103/PhysRevResearch.4.023217} {\bibfield  {journal} {\bibinfo  {journal} {Phys. Rev. Res.}\ }\textbf {\bibinfo {volume} {4}},\ \bibinfo {pages} {023217} (\bibinfo {year} {2022})}\BibitemShut {NoStop}%
\bibitem [{\citenamefont {Van~den Broeck}(2013)}]{van2013efficiency}%
  \BibitemOpen
  \bibfield  {author} {\bibinfo {author} {\bibfnamefont {C.}~\bibnamefont {Van~den Broeck}},\ }\bibfield  {title} {\bibinfo {title} {Efficiency at maximum power in the low-dissipation limit},\ }\href {https://doi.org/10.1209/0295-5075/101/10006} {\bibfield  {journal} {\bibinfo  {journal} {Europhys. Lett.}\ }\textbf {\bibinfo {volume} {101}},\ \bibinfo {pages} {10006} (\bibinfo {year} {2013})}\BibitemShut {NoStop}%
\bibitem [{\citenamefont {Ondrechen}\ \emph {et~al.}(1981)\citenamefont {Ondrechen}, \citenamefont {Andresen}, \citenamefont {Mozurkewich},\ and\ \citenamefont {Berry}}]{ondrechenMaximumWorkFinite1981}%
  \BibitemOpen
  \bibfield  {author} {\bibinfo {author} {\bibfnamefont {M.~J.}\ \bibnamefont {Ondrechen}}, \bibinfo {author} {\bibfnamefont {B.}~\bibnamefont {Andresen}}, \bibinfo {author} {\bibfnamefont {M.}~\bibnamefont {Mozurkewich}},\ and\ \bibinfo {author} {\bibfnamefont {R.~S.}\ \bibnamefont {Berry}},\ }\bibfield  {title} {\bibinfo {title} {Maximum work from a finite reservoir by sequential {{Carnot}} cycles},\ }\href {https://doi.org/10.1119/1.12426} {\bibfield  {journal} {\bibinfo  {journal} {Am. J. Phys.}\ }\textbf {\bibinfo {volume} {49}},\ \bibinfo {pages} {681} (\bibinfo {year} {1981})}\BibitemShut {NoStop}%
\bibitem [{\citenamefont {Ma}(2020)}]{maEffectFiniteSizeHeat2020}%
  \BibitemOpen
  \bibfield  {author} {\bibinfo {author} {\bibfnamefont {Y.-H.}\ \bibnamefont {Ma}},\ }\bibfield  {title} {\bibinfo {title} {Effect of {{Finite-Size Heat Source}}'s {{Heat Capacity}} on the {{Efficiency}} of {{Heat Engine}}},\ }\href {https://doi.org/10.3390/e22091002} {\bibfield  {journal} {\bibinfo  {journal} {Entropy}\ }\textbf {\bibinfo {volume} {22}},\ \bibinfo {pages} {1002} (\bibinfo {year} {2020})}\BibitemShut {NoStop}%
\bibitem [{\citenamefont {Chen}\ \emph {et~al.}(2021)\citenamefont {Chen}, \citenamefont {Sun},\ and\ \citenamefont {Dong}}]{chen2021extrapolating}%
  \BibitemOpen
  \bibfield  {author} {\bibinfo {author} {\bibfnamefont {J.-F.}\ \bibnamefont {Chen}}, \bibinfo {author} {\bibfnamefont {C.~P.}\ \bibnamefont {Sun}},\ and\ \bibinfo {author} {\bibfnamefont {H.}~\bibnamefont {Dong}},\ }\bibfield  {title} {\bibinfo {title} {Extrapolating the thermodynamic length with finite-time measurements},\ }\href {https://doi.org/10.1103/PhysRevE.104.034117} {\bibfield  {journal} {\bibinfo  {journal} {Phys. Rev. E}\ }\textbf {\bibinfo {volume} {104}},\ \bibinfo {pages} {034117} (\bibinfo {year} {2021})}\BibitemShut {NoStop}%
\end{thebibliography}%
\end{document}